\pdfoutput=1
\documentclass[11pt,a4paper]{iopart}
\usepackage{iopams}
\usepackage{amsopn}
\usepackage[pdftex]{graphicx}
\usepackage{bm,bbm}
\usepackage{subfigure}

\newcounter{shadowtheorems}
\newtheorem{proposition}[shadowtheorems]{Proposition}
\newtheorem{corollary}[shadowtheorems]{Corollary}
\newtheorem{lemma}[shadowtheorems]{Lemma}
\newcommand{\proof}{\noindent {\it Proof.\ }}
\newcommand{\proofof}[1]{\noindent\emph{Proof of #1.}}
\newcommand{\bra}[1]{\langle #1 |}
\newcommand{\ket}[1]{| #1 \rangle}

\newcommand{\halmos}{$\square$}

\begin{document}

\title[Numerical shadow and geometry of quantum states]%
    {Numerical shadow and geometry of quantum states}

\author{Charles~F.~Dunkl$^1$, 
Piotr Gawron$^2$,
John~A.~Holbrook$^3$, 
Jaros{\l}aw A. Miszczak$^2$,  
Zbigniew Pucha{\l}a$^2$ and 
Karol~{\.Z}yczkowski$^{4,5}$}

\address{$^1$ Department of Mathematics, University of Virginia, 
Charlottesville, VA 22904---4137, USA} 

\address{$^2$ Institute of Theoretical and Applied Informatics, Polish Academy
of Sciences, Ba{\l}tycka 5, 44-100 Gliwice, Poland} 

\address{$^3$ Department of Mathematics and Statistics, University of Guelph,
Guelph, Ontario, N1G 2W1, Canada}

\address{$^4$ Institute of Physics,  Jagiellonian University, 
 Reymonta 4, 30-059 Krak{\'o}w, Poland} 

\address{$^5$ Center for Theoretical Physics, Polish Academy of Sciences, 
         Aleja Lotnik{\'o}w 32/44, 02-668 Warszawa, Poland}

\ead{cfd5z@virginia.edu \quad gawron@iitis.pl \quad jholbroo@uoguelph.ca \quad
miszczak@iitis.pl \quad z.puchala@iitis.pl \quad karol@tatry.if.uj.edu.pl}

\begin{abstract}
The totality of normalised density matrices of order $N$ 
forms a convex set ${\cal Q}_N$ in ${\mathbbm R}^{N^2-1}$.
Working with the flat geometry induced by the Hilbert--Schmidt distance we
consider images of orthogonal projections of ${\cal Q}_N$ onto a two--plane and show
that they are similar to the numerical ranges of matrices of order
$N$. For a matrix $A$ of a order $N$ one defines its numerical shadow as a
probability distribution supported on its numerical range $W(A)$, induced by the
unitarily invariant Fubini--Study measure on the
complex projective manifold ${\mathbbm C}P^{N-1}$. We define generalized,
mixed--states shadows of $A$ and demonstrate their usefulness to analyse the
structure of the set of quantum states and unitary dynamics therein.

\end{abstract}

% matrix theory - 02.10.Yn
% quantum information - 03.67.-a
% operator theory - 02.30.Tb 	
\vspace{2pc}
\noindent\small{April 14, 2011}

% \submitto{\JPA}

\maketitle

%%%%%%%%%%%%%%%%%%%%%%%%%%%%%%%%%%%%%%%%%%%%%%%%%%%%%%%%%%%%%%%%%%%%%%%%%%%%%%%%
\section{Introduction}
%%%%%%%%%%%%%%%%%%%%%%%%%%%%%%%%%%%%%%%%%%%%%%%%%%%%%%%%%%%%%%%%%%%%%%%%%%%%%%%%
Investigation of the geometry of the set of quantum states remains a subject of
current scientific interests in view of possible applications in the theory of
quantum information processing. The set $\Omega_N$ of pure quantum states
belonging to a $N$ dimensional complex Hilbert space ${\cal H}_N$ is known to be
equivalent to the complex projective space, $\Omega_N = {\mathbbm C}P^{N-1}$, of
$2N-2$ real dimensions. However, as this set is embedded into the $N^2-1$
dimensional set ${\cal Q}_N$ of density matrices of size $N$ by a non-linear
constraint, $\rho=\rho^2$, the geometric structure of the set of mixed quantum
states is rather involved \cite{MW95,BZ06}. The only simple case corresponds to
the one--qubit system, $N=2$.

The set $\Omega_2$ of $N=2$ pure states forms the {\sl Bloch sphere}, ${\mathbbm
C}P^{1}=S^2$, with respect to the standard Hilbert-Schmidt metric. The $3$-disk
inside the sphere, often called the {\sl Bloch ball}, represents the set ${\cal
Q}_2$ of one--qubit mixed states. In this simple case any projection of this set
onto a plane forms an ellipse, which can be degenerated to an interval. In the
case of $N=3$ the $8$ dimensional set ${\cal Q}_3$ of one--qutrit mixed states
is neither a polytope nor an ellipsoid \cite{ZS01,Ki03,SZL04}, and the set
$\Omega_3={\mathbbm C}P^{2}$ of its extremal states 
is connected and has four real dimensions.

Due to the high dimensionality of the problem our understanding of the geometry
of the set ${\cal Q}_N$ of mixed states is still rather limited. This set forms
a convex body which contains an in--ball of radius $r_N=\sqrt{1/N(N-1)}$ and can
be inscribed into an out--sphere of radius $R_N=(N-1)r_N=\sqrt{(N-1)/N}$
\cite{BZ06}. Some information on the subject can be gained by studying the
$2$--dimensional cross-sections of ${\cal Q}_N$ as demonstrated in
\cite{JS01,VDM02,KK05} for $N=3$ and $N=4$. Another option is to investigate
projections of this set into a plane -- such an approach was advocated for $N=3$
in \cite{We10}. As the set ${\cal Q}_N$ of quantum states is convex, also its
cross-sections and projections inherit convexity.

In this work we study the general structure of a two--dimensional projection of
the set ${\cal Q}_N$ of mixed states. A bridge between the geometry of the set
of quantum states and the notion of numerical range used in operator theory is
established. For any operator $A$, acting on the complex Hilbert space ${\cal
H}_N$, one defines its {\sl numerical range} \cite{HJ2,GR97}
(also called {\sl field of values})
 as a subset of the complex plane which contains expectation
values of $A$ among arbitrary normalized pure states,
\begin{equation}
W(A)=\{ z: z=\langle \psi | A| \psi \rangle, \   |\psi\rangle \in {\cal H}_N, 
\  \langle \psi| \psi \rangle =1\}.
\label{range1}
\end{equation}

We analyse the set of orthogonal projections of the set ${\cal Q}_N$ onto a
$2$--plane and prove that it is equivalent to the set of all possible numerical
ranges of complex matrices of order $N$. Numerical ranges of normal matrices of
size $N$ correspond to orthogonal projections of the set ${\cal C}_N$ of
classical states -- the $(N-1)$-dimensional simplex $\Delta_{N-1}\subset
{\mathbb R}^{N-1}$.

Further information on the structure of the set of quantum states
of a size $N$ can be obtained by
 studying the {\sl numerical shadow}
\cite{Zy+10,DGHPZ11,GS10} of various matrices of order $N$.
For any operator $A$ acting on ${\cal H}_N$ one defines a probability distribution
$P_A(z)$ on the complex plane, supported in the numerical range $W(A)$,
\begin{equation}
P_A(z) := \int_{\Omega_N} {\rm d} \mu(\psi) 
 \delta\Bigl( z-\langle \psi|A|\psi\rangle\Bigr)  .
\label{shadow}  
\end{equation}
Here $\mu(\psi)$ denotes the unique unitarily invariant (Fubini-Study) measure
on the set $\Omega_N$ of $N$-dimensional pure quantum states. In other words the
shadow $P$ of matrix $A$ at a given point $z$ characterizes the likelihood that
the expectation value of $A$ among a random pure state is equal to $z$.

The distribution $P_A(z)$ is naturally associated a given matrix $A$, and some
of its properties were described in \cite{DGHPZ11}. In this work we advocate a
complementary approach and show that investigating the shadows of several
different complex matrices $A$ of a fixed size $N$ contributes to our
understanding of the structure of the entire set ${\cal Q}_N$ of quantum states.
In a sense, the choice of a matrix $A$ corresponds to the selection of the
plane, onto which the set of quantum states is projected.

This paper is organized as follows. In \sref{sec:basic-facts} we fix the
notation and introduce necessary concepts.
A link between $2$--dimensional projections of the set of quantum states of a
given size $N$ and the set of possible numerical ranges of matrices of order $N$
is presented in \sref{sec:range-projection}. 
In \sref{sec:classes} we analyze different classes of numerical shadows
of matrices of small order $N=2,3,4$.
Unitary dynamics of a pure quantum state in the background of numerical shadow
is presented in \sref{sec:dynamics}.
\Sref{sec:mixed-shadow} is devoted to the mixed--states numerical shadow,
which corresponds to a projection of the full set ${\cal Q}_N$ of density
matrices onto a plane.
The case of a large dimension, $N\gg 1$, is treated in \sref{sec:large-n}
jointly with the shadow of random matrices.
Finally, in \sref{sec:concluding} we provide some concluding remarks and
summarize the contribution of this paper.

%%%%%%%%%%%%%%%%%%%%%%%%%%%%%%%%%%%%%%%%%%%%%%%%%%%%%%%%%%%%%%%%%%%%%%%%%%%%%%%%
\section{Classical and quantum states}\label{sec:basic-facts} 
%%%%%%%%%%%%%%%%%%%%%%%%%%%%%%%%%%%%%%%%%%%%%%%%%%%%%%%%%%%%%%%%%%%%%%%%%%%%%%%%
Let $p=\{ x_1,x_2,\dots, x_N\}$ be a normalized probability vector, so $x_i\ge
0$ and $\sum_{i=1}^N x_i =1$. Such a vector represents a {\sl classical state},
and the set ${\cal C}_N$ of all classical states forms an $(N-1)$ dimensional
regular simplex $\Delta_{N-1} \subset {\mathbbm R}^{N-1}$. There exist exactly
$N$ {\sl classical pure states}, which correspond to the corners of the simplex.
All other classical states can be expressed by a convex combination of pure
states and are called {\sl mixed}. Typical mixed states are characterized by the
full rank and they form the entire interior of the probability simplex.

In quantum theory one describes a physical system with $N$ distinguishable
states by elements of a complex Hilbert space ${\cal H}_N$ of size $N$. Its
elements represent pure quantum states, $|\psi\rangle \in {\cal H}_N$. Quantum
states are assumed to be normalized, $||\psi||^2=\langle \psi| \psi\rangle=1$,
so they belong to the sphere of dimension $2N-1$. Since one identifies two
states, which differ by a global phase only, $|\psi\rangle \sim |\phi\rangle =
e^{- \alpha}|\psi\rangle$, the set of all pure quantum states $\Omega_N$, which
act on ${\cal H}_N$, is equivalent to the complex projective space,
$\Omega_N={\mathbbm C}P^{N-1}$ \cite{BZ06}.

In analogy to the classical case, one also defines
mixed quantum states (density matrices) by a convex combination
of projectors onto pure states,
$\rho=\sum_i p_i |\psi_i\rangle\langle \psi_i|$,
where $p_i>0$ and $\sum_i p_i=1$.
Let us denote the set of all density matrices 
of order $N$ by ${\cal Q}_N$.
It contains all density operators which are
positive and normalized,
\begin{equation}
{\cal Q}_N=\{\rho:{\cal H}_N \to {\cal H}_N, \quad
\rho^*=\rho, \quad
  \rho\ge 0, \quad {\rm Tr} \rho=1 \} \; .
\label{densN}
\end{equation}
Since density operators are Hermitian and normalized,
this set is $N^2-1$ dimensional. 
It includes the set of classical states, 
$ {\cal Q}_N \supset {\cal C}_N=\Delta_{N-1}$,
as well as the set of pure quantum states, 
${\cal Q}_N \supset \Omega_N={\mathbbm C}P^{N-1}$.
We are going to work with the geometry implied by the
Hilbert--Schmidt norm of a matrix,
$|A|_{\rm HS}:=
\sqrt{  \mathrm{Tr} (  A^{\ast }A)}$,
and the Hilbert-Schmidt distance in the space of matrices,
\begin{equation}
d_{\rm HS}(A,B):= |A-B|_{\rm HS} = \sqrt{\mathrm{Tr}(A-B)(A-B)^{\ast}} .
\label{hsdist}
\end{equation}
It will be also convenient to define a real inner-product
by setting the polar identity
\begin{equation}
\left\langle A,B\right\rangle  =\frac{1}{4}| A+B |
_{\rm HS}^{2}-\frac{1}{4}|A-B|_{\rm HS}^{2}
 =\frac{1}{2} [ \mathrm{Tr}(A^{\ast}B+ B^{\ast}A)] .
\label{realhs}
\end{equation}
If $A^{\ast}=A$ and $B^{\ast}=B$ then $\left\langle A,B\right\rangle
=\mathrm{tr}AB$.

In the set $\Omega_N$ of quantum pure states one defines the Fubini-Study
measure $\mu_{\rm FS}$, which is induced by the Haar measure on $U(N)$ and is
invariant with respect to unitary transformations. In the case of one--qubit
states this measure corresponds to the uniform distribution of points on the
Bloch sphere $S^2$.

In practice, to generate pure states at random
according to the  measure $\mu_{\rm FS}$ it is sufficient
to generate uniformly points at the sphere $S^{2N-1}$.
One may also select an arbitrary column, (or row)
of a random unitary matrix $U$ distributed according to the Haar
measure. It directly gives the set of $N$ coefficients of the
random state in a given basis, $|\psi\rangle=\sum_{i=1}^N c_i |i\rangle$.
For instance, choosing the first column of $U$ we set  $c_i=U_{i,1}$
for $i=1,\dots,N$. Alternatively, one may generate $N$ independent
complex random numbers $z_i$ and renormalise them,
 $c_i=z_i/\sqrt{\sum_i |z_i|^2}$,
to obtain the desired distribution \cite{De86,ZS01}.

In this work we are going to use the following

%{\bf Proposition 1}. {\sl 
\begin{proposition}
Let $|\psi\rangle \in \Omega_N$ be a random pure state of size $N$ distributed
according to the Fubini-Study measure. If one represents it in an arbitrary
fixed basis, $|\psi\rangle =\sum_{i=1}^N c_i |i\rangle$ then the squared
absolute values of the coefficients, $p_i=|c_i|^2$, form a probability vector
distributed \emph{uniformly} in the probability simplex $\Delta_{N-1}$.
\end{proposition}
%}

This is equivalent to the known statement (see e.g. \cite{BZ06}), that the only
constraint on the components of a single column of a random unitary matrix $U$
distributed according to the Haar measure is the normalization condition,
$P(U_{11}, \dots U_{N1}) \sim \delta\Bigl(1-\sum_{i=1}^N |U_{i1}|^2\Bigr)$. This
fact directly implies 

%{\bf Corollary  2}. {\sl 
\begin{corollary} 
For any quantum state $\rho$ define a classical state $p={\rm diag}(\rho)$, so
$p_i=\rho_{ii}$. Then the Fubini-Study measure on the set $\Omega_N$ of quantum
pure states induces by this mapping the {\emph uniform} measure in the classical
probability simplex $\Delta_{N-1}$.
\end{corollary}
%}

In the case of $N=2$ the Fubini-Study measure covers uniformly the Bloch sphere
$S^2$. Working with the standard polar coordinates, $(r,\theta, \varphi)$, we
write the element of the volume of the unit sphere as $dS=d \varphi \sin \theta
d\theta=d \varphi d(\cos \theta)$. The polar angle $\theta$ is defined with
respect to the axis $z$, so the projection of a point of the sphere at this axis
reads $z=\cos \theta$. Hence the Fubini-Study measure implies the uniform
distribution $d(\cos \theta)= dz $ along the one-dimensional set $\Delta_1$ of
$N=2$ classical states.

%%%%%%%%%%%%%%%%%%%%%%%%%%%%%%%%%%%%%%%%%%%%%%%%%%%%%%%%%%%%%%%%%%%%%%%%%%%%%%%%
\section{Numerical range as a projection of the set of quantum states}\label{sec:range-projection}
%%%%%%%%%%%%%%%%%%%%%%%%%%%%%%%%%%%%%%%%%%%%%%%%%%%%%%%%%%%%%%%%%%%%%%%%%%%%%%%%

The set $\Omega_N={\mathbbm C}P^{N-1}$ of pure states of size $N$ forms the set
of extremal points in ${\cal Q}_N$. Any mixed state $\rho\in {\cal Q}_N$ can be
thus decomposed into a convex mixture of projectors $|\psi\rangle \langle
\psi|$. The expectation value of an operator $A$ among a pure state reads
$\langle \psi|A|\psi\rangle = {\rm Tr} \rho A$. Taking into account the
convexity of $W(A)$, the standard definition (\ref{range1}) of the numerical
range of $A$ can be therefore rewritten as~\cite{GPMSZ10}
\begin{equation}
W(A)=\{ z: z= {\rm Tr} \rho A, \ \rho \in {\cal Q}_N \} .
\label{range2}
\end{equation}

This expression suggests a possible link between 
numerical range and the structure of the set the ${\cal Q}_N$.
Usually one studies numerical range $W(A)$ for a given $A$~\cite{GR97}.
Here we propose to fix the dimension $N$ and consider the set of all
possible numerical ranges of matrices $A$ of this size
to analyze the geometry of quantum states.
More precisely, we establish the following facts.

%{\bf Proposition 3}.{\sl 
\begin{proposition}\label{prop:classical-states-shadows}
Let ${\cal C}_N$ denote the set of classical states of size $N$, which forms the
regular simplex $\Delta_{N-1}$ in ${\mathbbm R}^{N-1}$. Then the set of similar
images of orthogonal projections of ${\cal C}_N$ on a $2$--plane is equivalent
to the set of all possible numerical ranges $W(A)$ of all normal matrices $A$
(such that $AA^*=A^*A$) of order~$N$.
\end{proposition}
%}

%{\bf Proposition 4}.{\sl
\begin{proposition}\label{prop:quantum-states-shadows}
Let ${\cal Q}_N$ denote the set of quantum states size $N$ embedded in
${\mathbbm R}^{N^2-1}$ with respect to Euclidean geometry induced by
Hilbert-Schmidt distance. Then the set of similar images of orthogonal
projections ${\cal Q}_N$ on a $2$-plane is equivalent to the set of all possible
numerical ranges $W(A)$ of (arbitrary) matrices $A$ of order $N$.
\end{proposition}
%}

To prove the above propositions we will need an abstract lemma concerning the
real inner-product Euclidean spaces.

%{\bf Lemma 5}. {\sl 
\begin{lemma}\label{lem:euclidean-ortonormal}
Suppose $u_{1},u_{2},v_{0}\in V$, where $V$ is a Euclidean vector space (with
inner product $\left\langle \cdot,\cdot\right\rangle $ and norm $\left\vert
x\right\vert =\left\langle x,x\right\rangle ^{1/2}$), $v_{0}\neq0$ and
$\dim\left(  \mathrm{span}\left\{  u_{1},u_{2},v_{0}\right\}  \right)  \geq2$.
Then there exist real numbers $\alpha>0,\gamma_{1},\gamma_{2}$ such that
the vectors
\begin{equation}
v_{1}:=\frac{1}{\alpha}\left(  u_{1}+\gamma_{1}v_{0}\right), \ \ \
v_{2}:=\frac{1}{\alpha}\left(  u_{2}+\gamma_{2}v_{0}\right)  
\label{affin}
\end{equation}
{\sl are normalized and orthogonal,} 
\begin{equation}
\left\vert v_{1}\right\vert ^{2}=1=\left\vert v_{2}\right\vert ^{2}, \ \ \  
\left\langle v_{1},v_{2}\right\rangle =0 \; .
\label{otronorm}
\end{equation}
\end{lemma}
%}
\proof
Let $u_{i}^{\prime}=u_{i}-\frac{\left\langle u_{i},v_{0}\right\rangle }
{|v_{0}|^{2}} v_{0},\ \ i=1,2$.
By hypothesis $|u_{1}^{\prime}|^{2}+ |u_{2}^{\prime}|^{2}>0$. 
For $i=1,2$ set
$ c_{i}:=| v_{0}| \gamma_{i}+
\frac{\left\langle u_{i},v_{0}\right\rangle }{| v_{0}|}$
so that $v_{i}=\frac{1}{\alpha}\left(  u_{i}^{\prime}+\frac{c_{i}
}{|v_{0}| }v_{0}\right)  $. 
The desired equations become%
\begin{equation}
| u_{1}^{\prime}|^{2}+c_{1}^{2}  =\alpha^{2},\ \ 
| u_{2}^{\prime}|^{2}+c_{2}^{2}   =\alpha^{2},\ \
\left\langle u_{1}^{\prime},u_{2}^{\prime}\right\rangle +c_{1}c_{2}  =0 .
\end{equation}
Eliminating coefficient $\alpha$ 
we arrive at a quadratic equation for $c_{1}^{2}$ or $c_{2}^{2}$. Set%
\[
d=\left( | u_{1}^{\prime}|^{2}- |u_{2}^{\prime}|^{2}\right)^{2}
+4\langle u_{1}^{\prime},u_{2}^{\prime}\rangle^{2},
\]
then%
\begin{eqnarray}
c_{1}^{2}  &  =\frac{1}{2}
  \left(| u_{2}^{\prime}|^{2}-|u_{1}^{\prime}|^{2}\right)  
+\frac{1}{2}\sqrt{d},  \label{c1} \\
c_{2}^{2}  &  =\frac{1}{2}\left( |u_{1}^{\prime}|^{2}- 
|u_{2}^{\prime}|^{2}\right)  
+\frac{1}{2}\sqrt{d}, \label{c2} \\
\mathrm{sign}\left(  c_{1}c_{2}\right)   &  =-\mathrm{sign}\left\langle
u_{1}^{\prime},u_{2}^{\prime}\right\rangle ,\\
\alpha &  =\left(  \frac{1}{2}\left( |u_{1}^{\prime}|^{2}
+|u_{2}^{\prime}|^{2}\right)  
+\frac{1}{2}\sqrt{d}\right)  ^{1/2}.
\end{eqnarray}

Recall $|u_{1}^{\prime}|^{2}+|u_{2}^{\prime}|^{2}>0$
by hypothesis thus $\alpha>0$. There are generally two
solutions differing only in the signs of $c_{1}$ and $c_{2}$. 
If $\langle u_{1}^{\prime},u_{2}^{\prime}\rangle =0$ then 
$\sqrt{d}=\left\vert | u_{1}^{\prime}|^{2}-|u_{2}^{\prime}|^{2}\right\vert $, 
and one of the three following cases
apply:
\newline 1. 
$|u_{1}^{\prime}| > |u_{2}^{\prime}| \geq 0, \ c_{1}=0, \ 
c_{2}=\pm\sqrt{|u_{1}^{\prime}|^{2} -|u_{2}^{\prime}|^{2}}, \
\alpha=|u_{1}^{\prime}|$;
\newline 2. 
$|u_{2}^{\prime} | > |u_{1}^{\prime}|\geq 0, \
c_{1}=\pm\sqrt{ |u_{2}^{\prime}|^{2} - |u_{1}^{\prime}|^2}, \
c_{2}=0, \ \alpha = |u_{2}^{\prime}|$;
\newline3. 
$|u_{1}^{\prime}| = |u_{2}^{\prime}| >0, \ c_{1}=0, \ c_{2}=0, \
\alpha=|u_{1}^{\prime}|$. \ \
$\square$
%\end{proof}

Note that the formulae  (\ref{c1}, \ref{c2}) for $c_1$ and $c_2$
allow us to obtain the constants $\gamma_1$ and $\gamma_2$,
 which enter eq. \eref{affin}.
The scaling factor $\alpha=1$ if and only if
 $\langle u_{1}^{\prime},u_{2}^{\prime}\rangle ^{2}
= (  1-|u_{1}^{\prime}|^{2})  ( 1-|u_{2}^{\prime}|^{2})$, 
$|u_{1}^{\prime}|^{2}\leq 1$ and $| u_{2}^{'}|^{2}\leq 1$.

This lemma implies the following 

%{\bf Corollary 6}.{\sl 
\begin{corollary}
\label{cor:isomorphic-affine}
Suppose $E\subset\left\{  x\in V:\left\langle x,v_{0}\right\rangle =1\right\}
$ and $u_{1},u_{2}\in V$ define a linear map $\Phi:E\rightarrow\mathbb{C}$ by
$x\mapsto\left\langle x,u_{1}\right\rangle +\mathrm{i}\left\langle
x,u_{2}\right\rangle $. Unless $u_{1},u_{2}\in\mathbb{R}v_{0}$ in which case
$\Phi$ is constant, the map $\Phi$ is isometrically isomorphic to an orthogonal
projection followed by a similarity transformation (dilation and translation).
\end{corollary}
%}
\proof 
By Lemma \ref{lem:euclidean-ortonormal} there exist orthonormal vectors 
$v_{i}=\frac{1}{\alpha}\left( u_{i}+\gamma_{i}v_{0}\right) $ 
for $i=1,2$ and $\alpha>0$. 
Let $V_{0}=\mathrm{span}\left\{  v_{1},v_{2}\right\} $.
The orthogonal projection onto
$V_{0}$ is given by $\pi x:=\left\langle x,v_{1}\right\rangle v_{1}
+\left\langle x,v_{2}\right\rangle v_{2}$ 
and this is the general form of a rank 2 orthogonal projection.
The linear map 
$\theta:a_{1}v_{1}+a_{2} v_{2}\mapsto a_{1}+\mathrm{i}a_{2}$ 
is an isometry $V_{0}\rightarrow
\mathbb{C}$. If $x\in E$ then $\left\langle v_{0},x\right\rangle =1$ 
and
\begin{equation}
\theta\left(  \alpha\pi x-\left(  \gamma_{1}v_{1}+\gamma_{2}v_{2}\right)
\right)     =\theta\sum_{i=1}^{2} ( \langle u_{i}+\gamma_{i}
v_{0},x \rangle -\gamma_{i})  v_{i}
  =\theta\sum_{i=1}^{2}\left\langle u_{i},x\right\rangle v_{i}=\phi x. 
\end{equation}\halmos

Now we are ready to prove the main result of this paper, namely Proposition
\ref{prop:classical-states-shadows} and Proposition~\ref{prop:quantum-states-shadows}.

%%%%%%%%%%%%%%%%%%%%%%%%%%%%%%%%%%%%%%%%%%%%%%%%%%%%%%%%%%%%%%%%%%%%%%%%%%%%%%%%
\subsection{Normal matrices}
%%%%%%%%%%%%%%%%%%%%%%%%%%%%%%%%%%%%%%%%%%%%%%%%%%%%%%%%%%%%%%%%%%%%%%%%%%%%%%%%
\proofof{Proposition~\ref{prop:classical-states-shadows}}
Let $A$ be a normal matrix of order $N$ with eigenvalues
$\left\{  \lambda_{1},\ldots,\lambda_{N}\right\}  $. With respect to an
orthonormal basis of eigenvectors of $A$ one has $\sum_{i,j=1}^{N}%
\overline{\psi_{i}}A_{ij}\psi_{j}=
\sum_{i=1}^{N}\lambda_{i}|\psi_{i}|^{2}$ 
and the numerical range $W_{A}$ is the image of
the simplex $\Delta_{N-1}:=\left\{  t\in\mathbb{R}^{N}:t_{i}\geq0~\forall
i,\sum_{i=1}^{N}t_{i}=1\right\}  $ under the map
\begin{equation}
\Phi   :t\longmapsto\sum_{i=1}^{N}t_{i}\operatorname{Re}\lambda
_{i}+\mathrm{i}\sum_{i=1}^{N}t_{i}\operatorname{Im}\lambda_{i} 
 \  =  \ \langle t,u_{1}\rangle +\mathrm{i}\langle t,u_{2} \rangle ,
\label{normalmap}
\end{equation}
where $t\in \Delta_{N-1}$, and $\left( u_{1}\right) _{i}=\operatorname{Re}
\lambda_{i},\left( u_{2}\right) _{i}=\operatorname{Im}\lambda_{i}$ for $1\leq
i\leq N$. If $A\neq c {\mathbbm 1}$ (multiple of the identity, the eigenvalues
are all equal) then Lemma \ref{lem:euclidean-ortonormal} and Corollary \ref{cor:isomorphic-affine} apply with $v_{0}=\left(
1,\ldots,1\right)$, which completes the proof of
Proposition~\ref{prop:classical-states-shadows}.
\halmos

%%%%%%%%%%%%%%%%%%%%%%%%%%%%%%%%%%%%%%%%%%%%%%%%%%%%%%%%%%%%%%%%%%%%%%%%%%%%%%%%
\subsection{Non--normal matrices}
%%%%%%%%%%%%%%%%%%%%%%%%%%%%%%%%%%%%%%%%%%%%%%%%%%%%%%%%%%%%%%%%%%%%%%%%%%%%%%%%
\proofof{Proposition~\ref{prop:quantum-states-shadows}} The set ${\cal Q}_N$ of
quantum states (\ref{densN}) contains Hermitian operators $\rho$ which can be
diagonalized, $\rho=UDU^{\ast}$. Here $U$ is unitary while $D$ is a diagonal
matrix with $d_{ii}\geq0$ and $\sum_{i=1}^{N}d_{ii}=1$.

Consider any matrix $A$ of order $N$ and
write $\mathrm{Tr}\rho A=\mathrm{Tr}\rho A_{1}+\mathrm{i}\, \mathrm{Tr}\rho A_{2}$ with
$A_{1}=\frac{1}{2}\left(  A+A^{\ast}\right)$ and 
$A_{2}=\frac{1}{2\mathrm{i}}\left(  A-A^{\ast}\right)$.
 Lemma 5 and Corollary 6 apply now to the map
\begin{equation}
\Phi:\rho \longmapsto\mathrm{Tr}\rho A_{1}+\mathrm{i}\, \mathrm{Tr}\rho A_{2}
\label{nonmap}
\end{equation}
of the set ${\cal Q}_N$ onto numerical range $W(A)$ with 
$V$ %=M_{N}\left(  \mathbb{C}\right)  $
representing the linear space of complex matrices of size $N$
(or the real subspace of Hermitian matrices),
the real inner product  (\ref{realhs}),
and $v_{0}=I$, $u_{1}=A_{1}$, $u_{2}=A_{2}$ provided $A\neq c {\mathbbm 1}$.

Thus we have shown that for any matrix $A$ its numerical range
$W(A)$ is equal to an orthogonal projection of the set of density matrices.
To show the converse we may read formulae (\ref{affin}) backwards:
the projection of ${\cal Q}_N$ is determined by two orthonormal  Hermitian matrices 
$V_1$ and $V_2$, which then satisfy $|V_1|_{\rm HS}=|V_2|_{\rm HS}=1$
and $\mathrm{Tr}(V_1 V_2)=0$. Set $A = V_1 + \mathrm{i} V_2$,
 which gives now the required matrix
such that $W(A)$ is equal to the desired projection.
In this way a link between numerical ranges of generic matrices of order $N$
and projections of the set ${\cal Q}_N$ onto a two--plane
is established and  Proposition \ref{prop:quantum-states-shadows} is proved. $\square$

To obtain explicit formulae for the similarity transformation 
corresponding to an arbitrary matrix $A$ of order $N$ 
define three traceless matrices 
\begin{equation}
B  =A-\frac{\mathrm{Tr}A}{N}I,\ \
B_{1}  =\frac{1}{2}\left(  B+B^{\ast}\right) ,\ \
B_{2}  =\frac{1}{2\mathrm{i}}\left(  B-B^{\ast}\right)  .
\label{BBB}
\end{equation}
The latter two represent vectors in the Hilbert--Schmidt space 
and correspond to $u_{1}^{\prime}, u_{2}^{\prime}$ in Lemma 5.
Making use of the Hilbert--Schmidt norm we compute
the required coefficients for a given traceless matrix $B$
\begin{eqnarray}
\!\!\!\!\!\!\!\!\!\!\!\!\!\!\!\!
d & 
 =\mathrm{Tr}B^{2}\; \mathrm{Tr}B^{\ast2}= |\mathrm{Tr}B^{2}| ^{2},\ \ \ \
\alpha
  = \left(  \frac{1}{2} \mathrm{Tr}  (  BB^{\ast} )  
+\frac{1}{2} |\mathrm{Tr}B^{2}| \right)  ^{1/2},
\label{alpha}
\\
\!\!\!\!\!\!\!\!\!\!\!\!\!\!\!\!
c_{1}^{2} 
&  =-\frac{1}{4}\left(  \mathrm{Tr}B^{2}+\mathrm{Tr}B^{\ast2}\right)
+\frac{1}{2}| \mathrm{Tr}B^{2}| ,\ \ \ \
c_{2}^{2} 
 =\frac{1}{4}\left(  \mathrm{Tr}B^{2}+\mathrm{tr}B^{\ast2}\right)
+\frac{1}{2}| \mathrm{Tr}B^{2}| ,
\end{eqnarray}
and $\mathrm{sign}\left(  c_{1}c_{2}\right)  =-\mathrm{sign}
\langle u_{1}^{\prime},u_{2}^{\prime}\rangle =-\mathrm{sign}\left(
\operatorname{Im}\mathrm{Tr}B^{2}\right)  $.

%%%%%%%%%%%%%%%%%%%%%%%%%%%%%%%%%%%%%%%%%%%%%%%%%%%%%%%%%%%%%%%%%%%%%%%%%%%%%%%%
\section{Numerical shadow and quantum states}\label{sec:classes}
%%%%%%%%%%%%%%%%%%%%%%%%%%%%%%%%%%%%%%%%%%%%%%%%%%%%%%%%%%%%%%%%%%%%%%%%%%%%%%%%

The projectors $|\psi\rangle \langle \psi|$ onto pure states 
form extremal points of the set ${\cal Q}_N$ of quantum states,
hence the shape of a projection of the set $\Omega_N$ of pure states
onto a given plane coincides with the shape of the 
projection of the set of density matrices on  the same plane.
As shown in the previous section this set is equal to 
the numerical range $W(A)$ of a matrix $A$ of size $N$,
which determines the projection.

However, the differences appear if one studies not
only the support of the projection but also the 
corresponding probability measure.
A measure $P_A(z)$ determined by the
numerical shadow (\ref{shadow})
is induced by the Fubini--Study measure on the set 
$\Omega_N$ of pure state. Thus the standard numerical 
shadows of various matrices of size $N$
can be interpreted as a projection of the 
complex projective space, $\Omega_N={\mathbbm C}P^{N-1}$,
onto a plane. 
Before discussing in detail the cases of low dimensions,
let us present here some basic properties of the numerical shadow \cite{DGHPZ11}
(also called the {\sl numerical measure} \cite{GS10}). 

{
\renewcommand{\theenumi}{\textbf{\arabic{enumi}.}}
\renewcommand{\labelenumi}{\theenumi}
\begin{enumerate}
    \item By construction the distribution $P_A(z)$ is supported on the
    numerical range of $W(A)$ and it is normalized, $\int_{W(A)} P_A(z) d^2 z
    =1$.
    \item The (numerical) shadow is unitarily invariant, $P_A(z)=P_{UAU^*}(z)$.
    This is a consequence of the fact that the integration measure ${\rm d}
    \mu(\psi)$ is unitarily invariant.
    \item For any normal operator $A$ acting on ${\cal H}_N$, such that
    $AA^*=A^*A$, its shadow covers the numerical range $W(A)$ with the
    probability corresponding to a projection of a {\emph {regular}}
    $N$--simplex of classical states ${\cal C}_N$ (embedded in ${\mathbbm
    R}^{N-1}$) onto a plane.
    \item For a non--normal operator $A$ acting on ${\cal H}_N$, its shadow
    covers the numerical range $W(A)$ with the probability corresponding to an
    orthogonal projection of the complex projective manifold $\Omega_N={\mathbbm
    C}P^{N-1}$ onto a plane.
    \item For any two operators $A$ and $B$ acting on ${\cal H}_N$, the shadow
    of their tensor product does not depend on the order, \begin{equation} P_{A
    \otimes B}(z)=P_{B \otimes A} (z)\; . \label{prod} \end{equation} To show
    this property define a unitary swap operator $S$ which acts on a composite
    Hilbert space and interchanges the order in the tensor product, $S(|z\rangle
    \otimes |y\rangle)=|y\rangle \otimes |z\rangle$. Thus $ \langle x| A \otimes
    B |x \rangle = \langle x| S^{*} B \otimes A S |x\rangle$, and since $S$ is
    unitary it does not influence the numerical shadow induced by the unitarily
    invariant Fubini-Study measure on complex projective space.
\end{enumerate}
}

%%%%%%%%%%%%%%%%%%%%%%%%%%%%%%%%%%%%%%%%%%%%%%%%%%%%%%%%%%%%%%%%%%%%%%%%%%%%%%%%
\subsection{One--qubit states, $N=2$}
%%%%%%%%%%%%%%%%%%%%%%%%%%%%%%%%%%%%%%%%%%%%%%%%%%%%%%%%%%%%%%%%%%%%%%%%%%%%%%%%
Analysis of the numerical shadow is particularly simple in the case of matrices
of order $N=2$. The spectrum of the operator $A$ consists of two complex
numbers, $\sigma(A)=\{\lambda_1,\lambda_2\}$.

In the case of a normal matrix $A$ the numerical range $W(A)$ forms the closed
interval $[\lambda_1, \lambda_2]$, and the numerical shadow $P_A(z)$ covers this
interval uniformly~\cite{DGHPZ11}. 

If the matrix $A$ is non--normal the numerical range forms an elliptical disk
with $\lambda_1,\lambda_2$ as focal points and minor axis, $d=\sqrt{{\rm
Tr}AA^* - |\lambda_1|^2-|\lambda_2|^2}$. For a simple proof of this 1932
result of Murnaghan \cite{Mu32} see the note by Li \cite{Li96}.
In this generic case the numerical shadow 
is given by the probability distribution 
obtained by the projection of the hollow Bloch sphere 
of one--qubit pure states onto a plane \cite{DGHPZ11}.
In particular, the cross--section of the numerical shadow 
supported in an interval $x\in [0,1]$
is given by the arcsine distribution, 
$P(x) = 1/(\pi \sqrt{x(1-x)})$. Non-normal case is
shown in Fig.\ref{fig:shad2}, obtained for a matrix,
$$ 
 A^{(2)}_0 \ =   \  a_0 \left[\begin{array}{cc}
                1  & 1 \\
                0  & -1
\end{array}\right] \ .
%\quad \quad \quad 
% A^{(2)}_1 \ = \   a_1\left[\begin{array}{cc}
%                1  & 0 \\
%                0  & -1
%\end{array}\right] \ .
$$
For simplicity we have selected the centred matrix
such that ${\rm Tr}A=0$ so that one has $B=A$ in eq. (\ref{BBB}). 
The normalization constant
$a_0=\sqrt{2/5}$ 
is chosen in such a way that
the scaling constant defining the projection in (\ref{alpha})
is set to unity, $\alpha=1$, so the 
shadow of the set of quantum states is shown in its 'natural size':
The distance between both eigenvalues, $l=2a_1=\sqrt{2}$,
is equal to the diameter of the Bloch ball, $2R_2=\sqrt{2}$.

\newlength{\picwidth}
\setlength{\picwidth}{0.29\textwidth}
\begin{figure}[ht!]
\begin{center}
\subfigure[Shadow of matrix $A^{(2)}_0$]{\includegraphics[width=\picwidth]{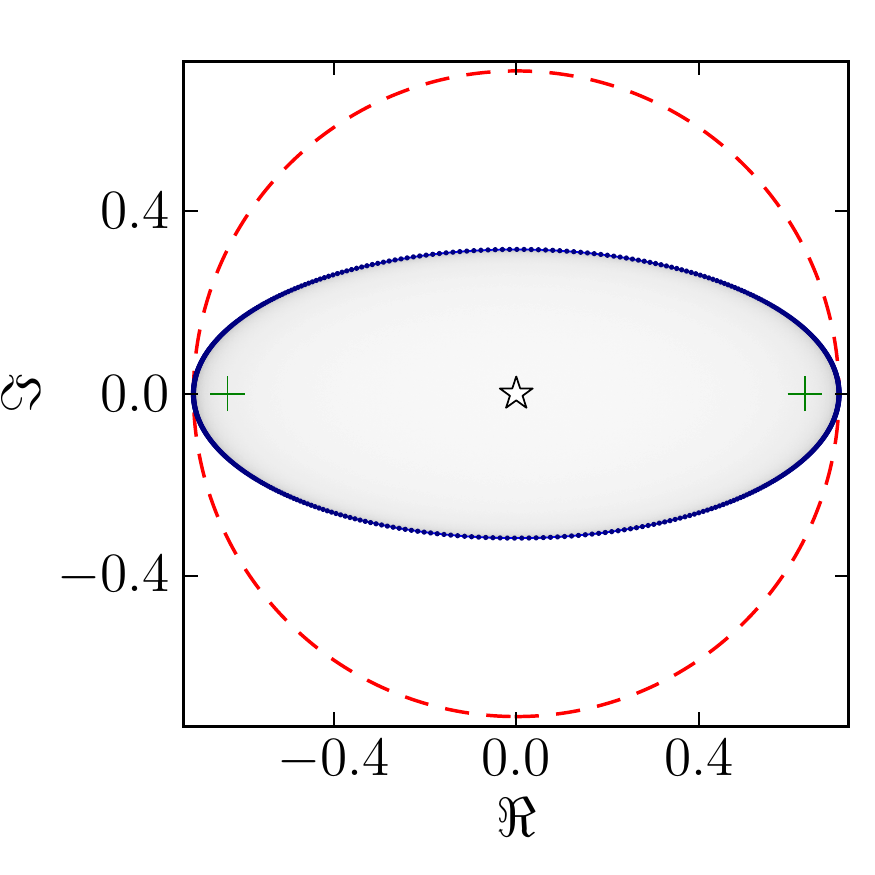}}
\subfigure[Cross-section for $\Im=0$]{\includegraphics[width=\picwidth]{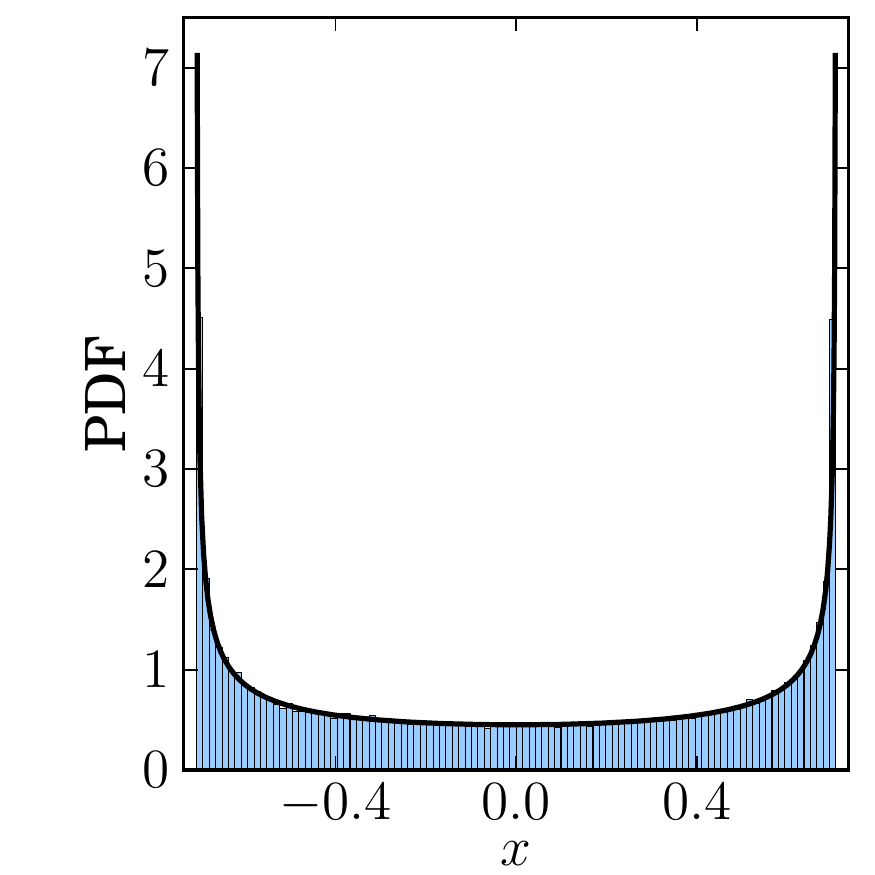}}
\caption{Projection of the set $\Omega_2$ of one--qubit
states generated by the numerical shadows of operators of order $N=2$
a) numerical shadow of generic matrix $A^{(2)}_0$ with an elliptical support.
Eigenvalues are denoted with crosses and dashed circle 
of radius $R_2=\sqrt{2}/2$ denotes the diameter of the Bloch ball.
Numerically obtained histogram is plotted in black, analytical plot is blue.
Plot is done for matrix translated in such a way that its trace
$(\star)$ is equal to zero and suitably rescaled.
b) Histogram of cross-section of the shadow supported in the interval $\left[-\frac{1}{\sqrt{2}},\frac{1}{\sqrt{2}}\right],$
solid line represents a probability density function
of the arcsine distribution 
$P(x)=\left(\pi \sqrt{
\frac{1}{2}-x^2
}
\right)^{-1}.$
}
\label{fig:shad2}
\end{center}
\end{figure}

%%%%%%%%%%%%%%%%%%%%%%%%%%%%%%%%%%%%%%%%%%%%%%%%%%%%%%%%%%%%%%%%%%%%%%%%%%%%%%%%
\subsection{One--qutrit states, $N=3$}
%%%%%%%%%%%%%%%%%%%%%%%%%%%%%%%%%%%%%%%%%%%%%%%%%%%%%%%%%%%%%%%%%%%%%%%%%%%%%%%%

The structure of the numerical range for 
$N=3$ was analyzed in detail by Keeler et al. \cite{KRS97}.
Numerical range of a matrix $A$ of order $N=3$ with spectrum
$\lambda_1, \lambda_2, \lambda_3$ forms:

{
\renewcommand{\theenumi}{\textbf{\alph {enumi})}}
\renewcommand{\labelenumi}{\theenumi}
\begin{enumerate}
    \item a compact set of an 'ovular' shape with three eigenvalues in its
    interior;
    \item a compact set with {\sl one} flat part -- e.g. the convex hull of a
    cardioid;
    \item a compact set with {\sl two} flat parts -- e.g. the convex hull of an
    ellipse and a point outside it;
    \item triangle. For any {\bf normal} matrix $A$ its numerical range is equal
    to the triangle spanned be the spectrum, $W(A)=\Delta(\lambda_1, \lambda_2,
    \lambda_3)$. In the latter case the numerical shadow can be verbally
    interpreted as the shadow of the set ${\cal C}_3$ of $N=3$ classical states
    -- a uniformly covered equilateral triangle $\Delta_2$.   
\end{enumerate}
}

\begin{figure}[ht!]
\begin{center}
\subfigure[$A^{(3)}_0$]{\includegraphics[width=\picwidth]{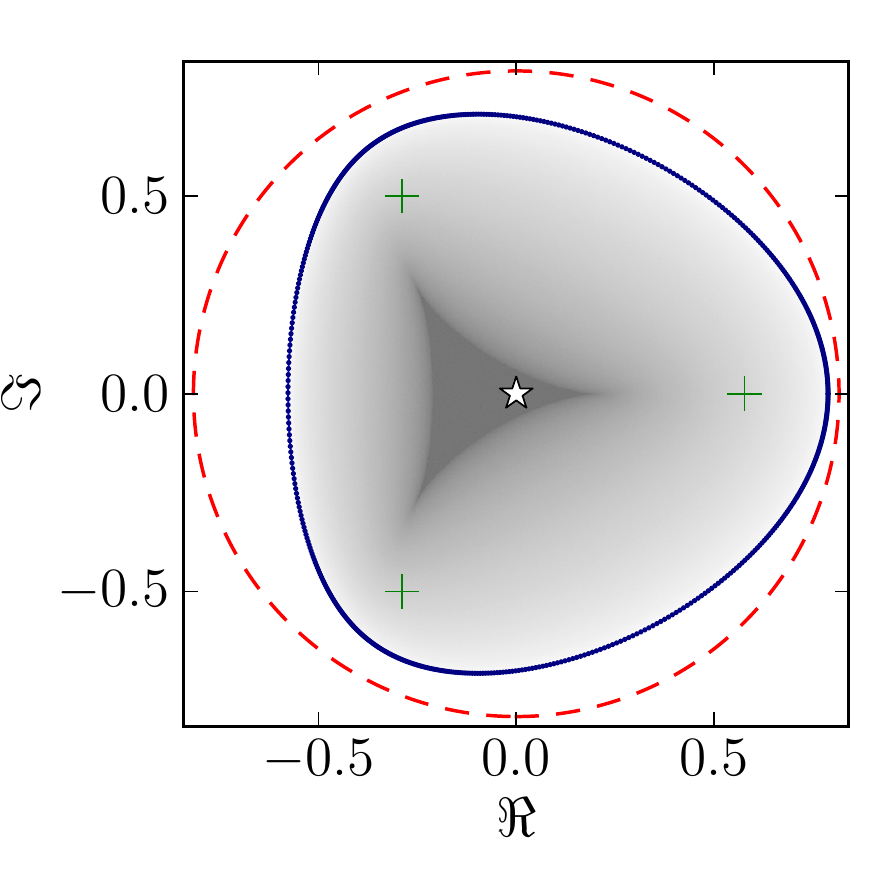}}
\subfigure[$A^{(3)}_1$]{\includegraphics[width=\picwidth]{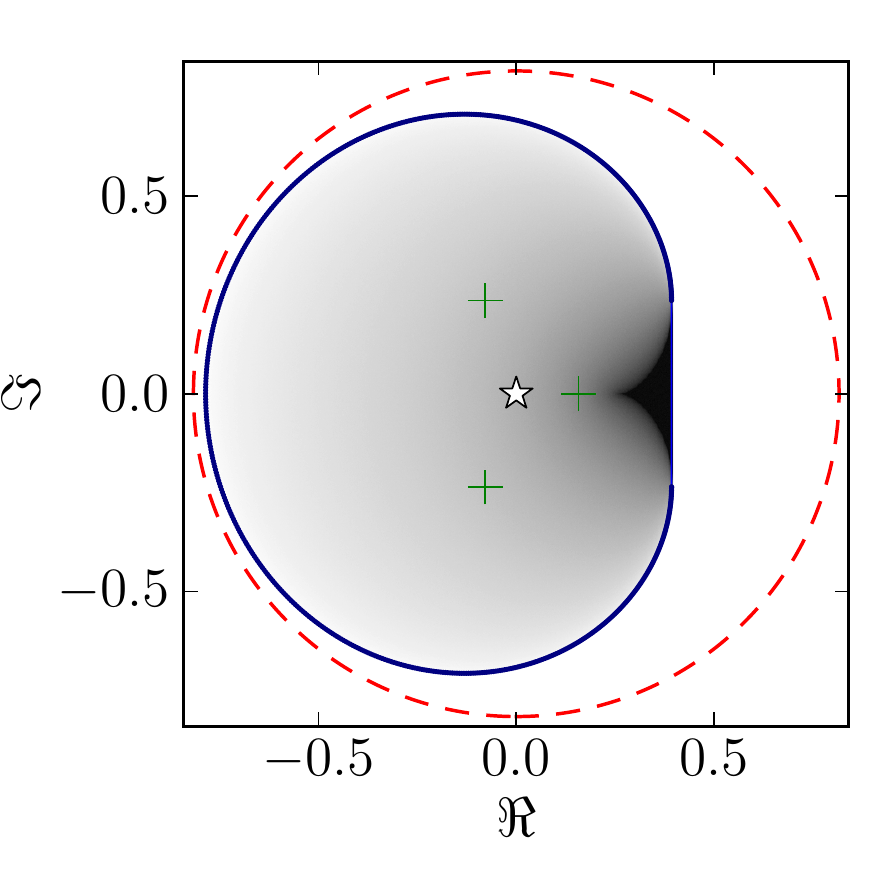}}\\
\subfigure[$A^{(3)}_2$]{\includegraphics[width=\picwidth]{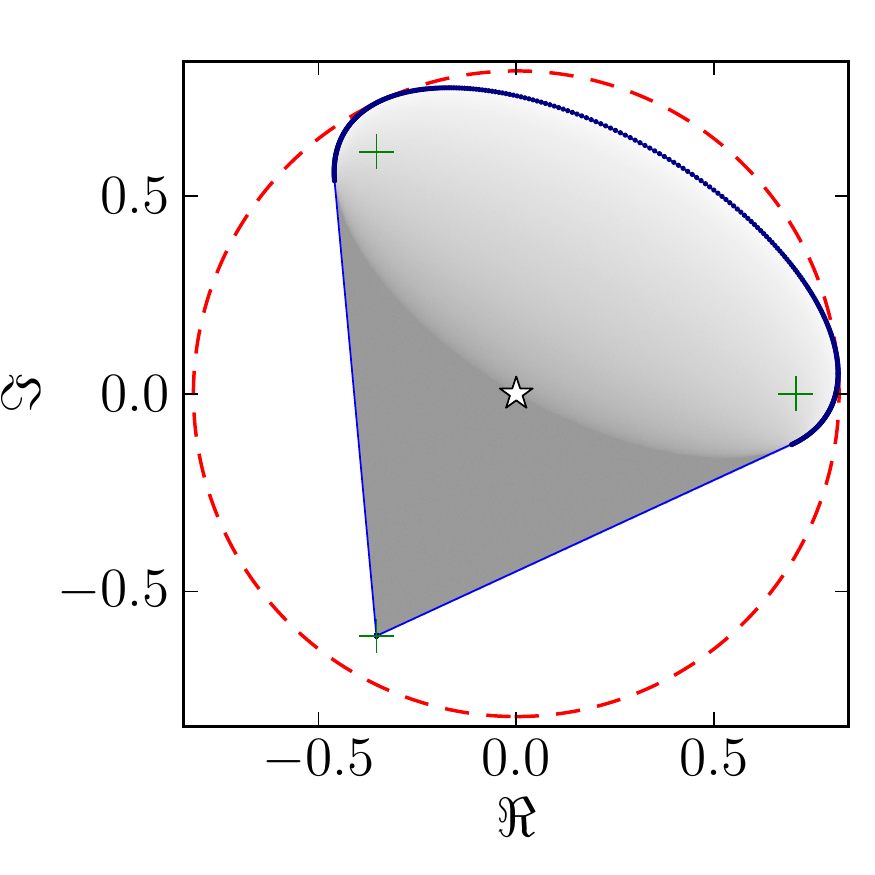}}
\subfigure[$A^{(3)}_3$]{\includegraphics[width=\picwidth]{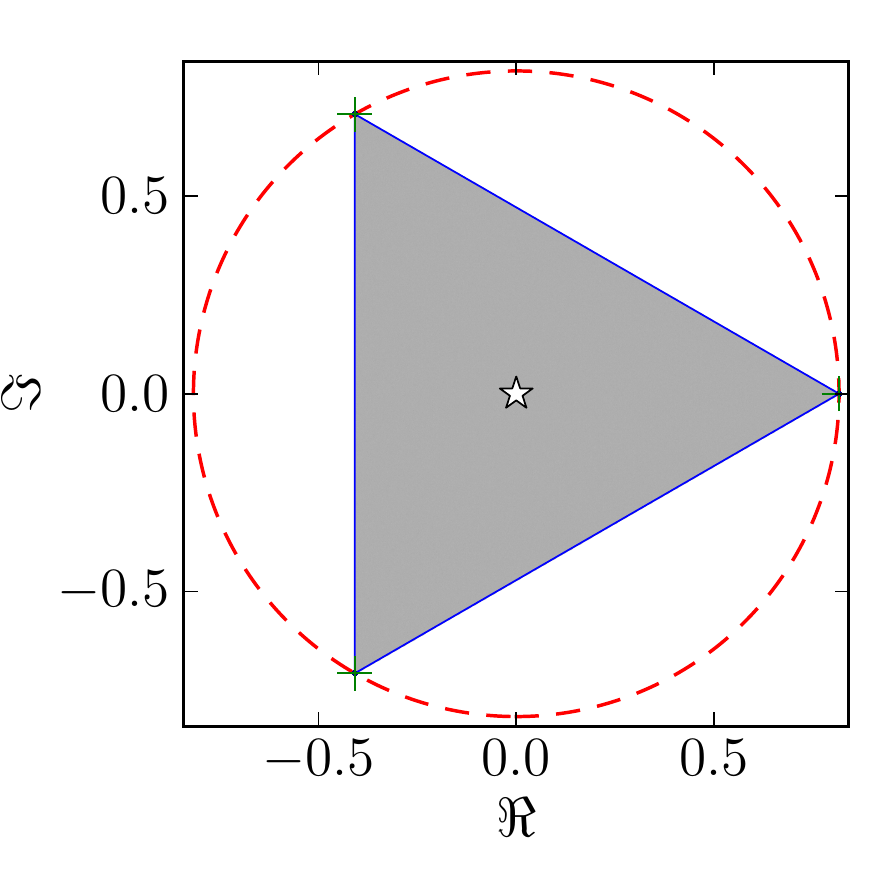}}
\caption{Projections of the set $\Omega_3$ of one--qutrit
states generated by the numerical shadows of operators of order $N=3$;
a) a generic matrix $A^{(3)}_0$ with an oval--like numerical shadow,
b) $A^{(3)}_1$ with one flat part of the boundary $\partial W$ of the numerical range,
c) $A^{(3)}_2$ a simple sum  with two flat parts of $\partial W$,
d) a diagonal normal matrix $A^{(3)}_3$ with numerical range
  equal to the triangle of eigenvalues, represented by $(+)$.
Dashed circle of radius $R_3$ represents projection of the sphere
in which  $\Omega_3$ is inscribed.
All plots are done for matrices translated in such a way that their trace
$(\star)$ is equal to zero and suitably rescaled.
}
\label{fig:shad3}
\end{center}
\end{figure}

The four classes of $N=3$ numerical ranges are illustrated in \fref{fig:shad3}.
It shows the numerical shadow supported on the corresponding numerical range,
obtained for
\begin{eqnarray*}
 A^{(3)}_0  =   a_0 \left[\begin{array}{ccc}
                1  & 1   & 1 \\
                0  & \omega_3 & 1 \\
                0  & 0   & \omega_3^2  
\end{array}\right]  ,
\  
A^{(3)}_1  =   a_1 
\left[
\begin{array}{ccc}
 5-3 \mathrm{i} & 0 & 6 \\
 0 & 5+3 \mathrm{i} & 6 \\
 -6 & -6 & -10
\end{array}
\right],
 \ \\
A^{(3)}_2  =   a_2 \left[\begin{array}{ccc}
                1  & 1   & 0 \\
                0  & \omega_3 & 0 \\
                0  & 0   & \omega_3^2  
\end{array}\right]  ,
\  
A^{(3)}_3  =   a_3 \left[\begin{array}{ccc}
                1  & 0   & 0 \\
                0  & \omega_3 & 0 \\
                0  & 0   & \omega_3^2  
\end{array}\right].
\end{eqnarray*}
The symbol $\omega_k$ denotes the $k$-th root of unity, so $\omega_3=\exp(\mathrm{i} 2\pi
/3)$. As before the matrices are chosen to be traceless, so $B=A$ in (\ref{BBB})
and the shadows are centered. Furthermore, the normalization constants are
designed to assure that the scaling constant in eq. (\ref{alpha}) in every case
is set to unity, $\alpha=1$, so the Figure shows images of the set of quantum
states in its natural size. For instance in the case of the diagonal matrix
$A^{(3)}_3$ the prefactor reads $a_3=\sqrt{2/3}$, so that the eigenvalues are
located at the distance $\sqrt{2/3}$ from the origin. This is just the radius
$R_3$ of the sphere in which the set ${\cal Q}_3$ is inscribed.

The study of the geometry of the numerical range was initiated by Kippenhahn
\cite{Ki51} and later developed by Fiedler \cite{Fi81}. In more recent papers
\cite{JAG98,He10} the differential topology and projection aspects of numerical
range were investigated. In particular it was shown \cite{JAG98} that the
numerical range of a generic matrix $A$ of order three pertains to the class a)
above, as the boundary of $W(A)$ does not contain intervals. Critical lines
inside the range, analysed in \cite{JAG98,He10} where shown to influence the
structure of the numerical shadow \cite{DGHPZ11}. Thus we may now relate the
critical lines with the geometry of complex projective spaces projected onto a
plane.

In the one--qutrit case $N=3$ obtained probability distributions can be
interpreted as images of the set of pure states $\Omega_3={\mathbbm C}P^{2}$ on
the plane. Although it is not so simple to imagine the structure of the complex
projective space \cite{BBZ02}, some experience is gained by studying numerical
shadows of various non-normal matrices of size $3$.

%%%%%%%%%%%%%%%%%%%%%%%%%%%%%%%%%%%%%%%%%%%%%%%%%%%%%%%%%%%%%%%%%%%%%%%%%%%%%%%%
\subsection{Four--level systems, $N=4$}
%%%%%%%%%%%%%%%%%%%%%%%%%%%%%%%%%%%%%%%%%%%%%%%%%%%%%%%%%%%%%%%%%%%%%%%%%%%%%%%%
Various shapes of the numerical range for matrices of size $N=4$ 
correspond to various projections of the set ${\cal Q}_4$ 
of quantum states of size four. As in the
case of the qutrit we analyse numerical shadows of traceless matrices normalized
such that the scaling constant $\alpha$ is set to unity.

Even though several results on geometry of numerical range for $N=4$
are available \cite{Gau06,LL09},
complete classification of numerical ranges in this case is still missing. 
To provide an overview of the possible structure of the numerical shadow 
we analysed the following matrices of order four,

{\footnotesize
\begin{eqnarray*}
A^{(4)}_0=\left[
\begin{array}{cccc}
 1 & 1 & 1 & 1 \\
 0 & i & 1 & 1 \\
 0 & 0 & -1 & 1 \\
 0 & 0 & 0 & -\mathrm{i}
\end{array}
\right],
&
A^{(4)}_1=\left[
\begin{array}{cccc}
 \mathrm{i} & 0 & -1 & 0 \\
 0 & 0 & -1 & 0 \\
 1 & 1 & 1-\mathrm{i} & 0 \\
 0 & 0 & 1 & 1
\end{array}
\right],
&
A^{(4)}_2=\left[
\begin{array}{cccc}
 1 & 0 & 0 & 1 \\
 0 & \mathrm{i} & 0 & 1 \\
 0 & 0 & -1 & 0 \\
 0 & 0 & 0 & -\mathrm{i}
\end{array}
\right], \\
A^{(4)}_3=\left[
\begin{array}{cccc}
 1 & 0 & 0 & 1 \\
 0 & \mathrm{i} & 1 & 0 \\
 0 & 0 & -1 & 0 \\
 0 & 0 & 0 & -\mathrm{i}
\end{array}
\right],
&
A^{(4)}_4=\left[
\begin{array}{cccc}
 1 & 0 & 0 & 1 \\
 0 & \mathrm{i} & 0 & 0 \\
 0 & 0 & -1 & 0 \\
 0 & 0 & 0 & -\mathrm{i}
\end{array}
\right],
&
A^{(4)}_5=\left[
\begin{array}{cccc}
 \mathrm{i} & 0 & -1 & 0 \\
 0 & 0 & -1 & 0 \\
 1 & 1 & 1-\mathrm{i} & 0 \\
 0 & 0 & 0 & 1
\end{array}
\right], \\
A^{(4)}_6=\left[
\begin{array}{cccc}
 1 & 0 & 1 & 0 \\
 0 & \mathrm{i} & 0 & 1 \\
 0 & 0 & -1 & 0 \\
 0 & 0 & 0 & -\mathrm{i}
\end{array}
\right],
&
A^{(4)}_7=\left[
\begin{array}{cccc}
 1 & 0 & 0 & 0 \\
 0 & \mathrm{i} & 0 & 1 \\
 0 & 0 & -1 & 0 \\
 0 & 0 & 0 & -\mathrm{i}
\end{array}
\right],
&
A^{(4)}_8=\left[
\begin{array}{cccc}
 1 & 0 & 0 & 0 \\
 0 & \mathrm{i} & 0 & 0 \\
 0 & 0 & -1 & 0 \\
 0 & 0 & 0 & -\mathrm{i}
\end{array}
\right].
\end{eqnarray*}
}

\begin{figure}[h!]
\begin{center}
\subfigure[$A^{(4)}_0$]{\includegraphics[width=\picwidth]{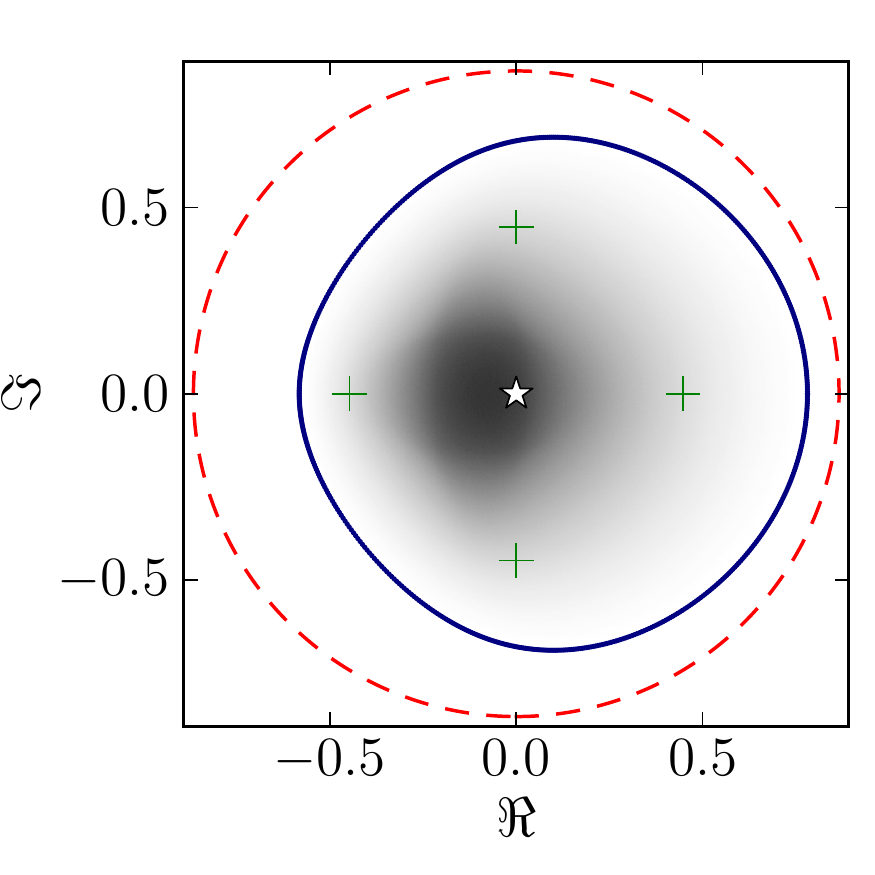}}
\subfigure[$A^{(4)}_1$]{\includegraphics[width=\picwidth]{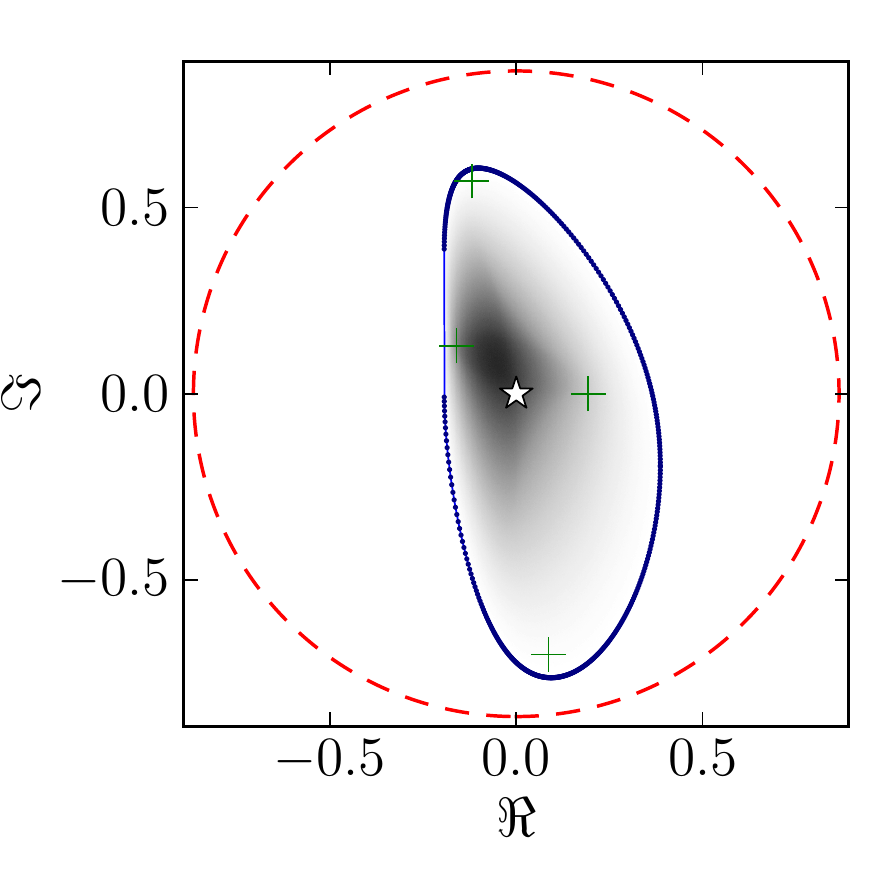}}
\subfigure[$A^{(4)}_2$]{\includegraphics[width=\picwidth]{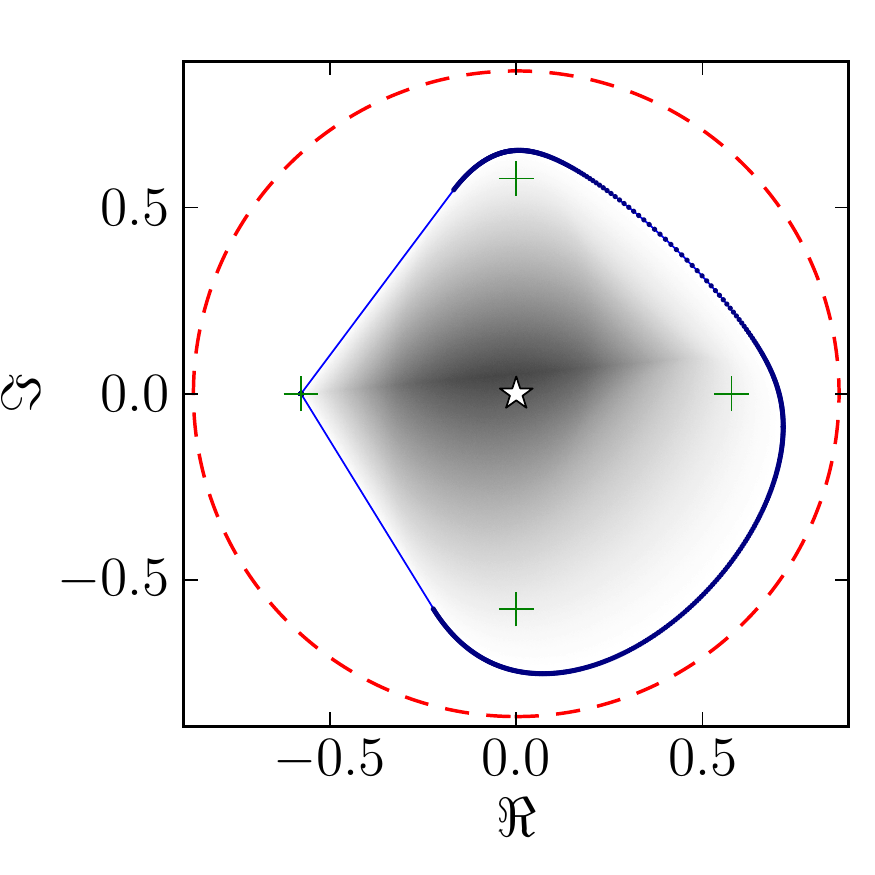}}
\subfigure[$A^{(4)}_3$]{\includegraphics[width=\picwidth]{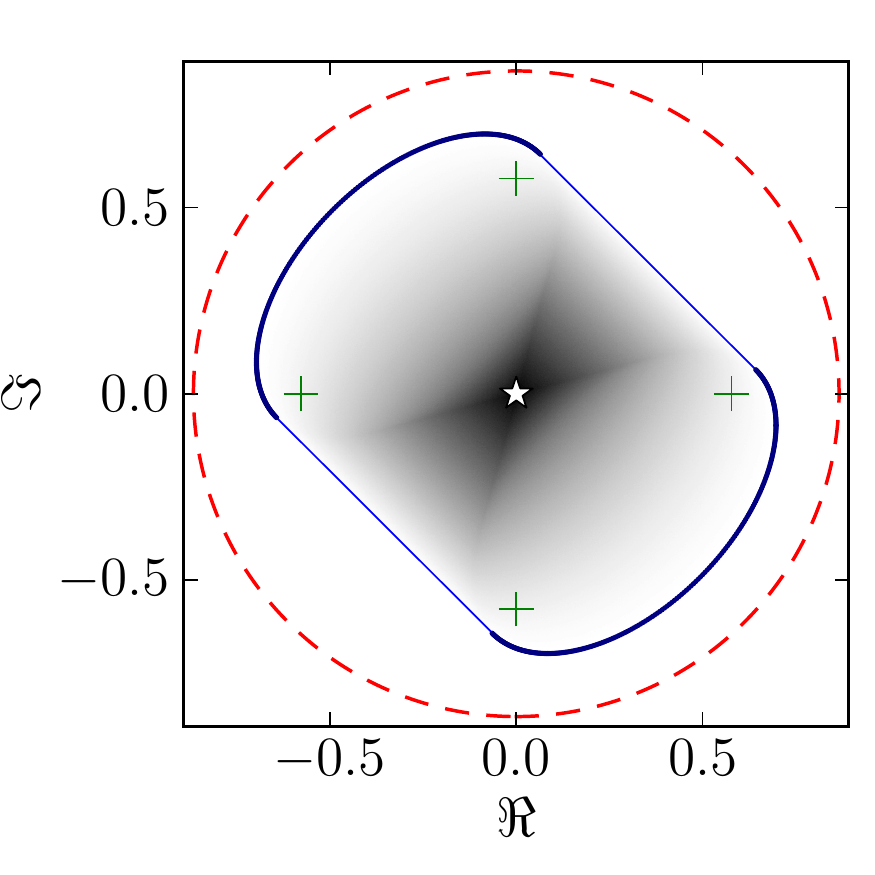}}
\subfigure[$A^{(4)}_4$]{\includegraphics[width=\picwidth]{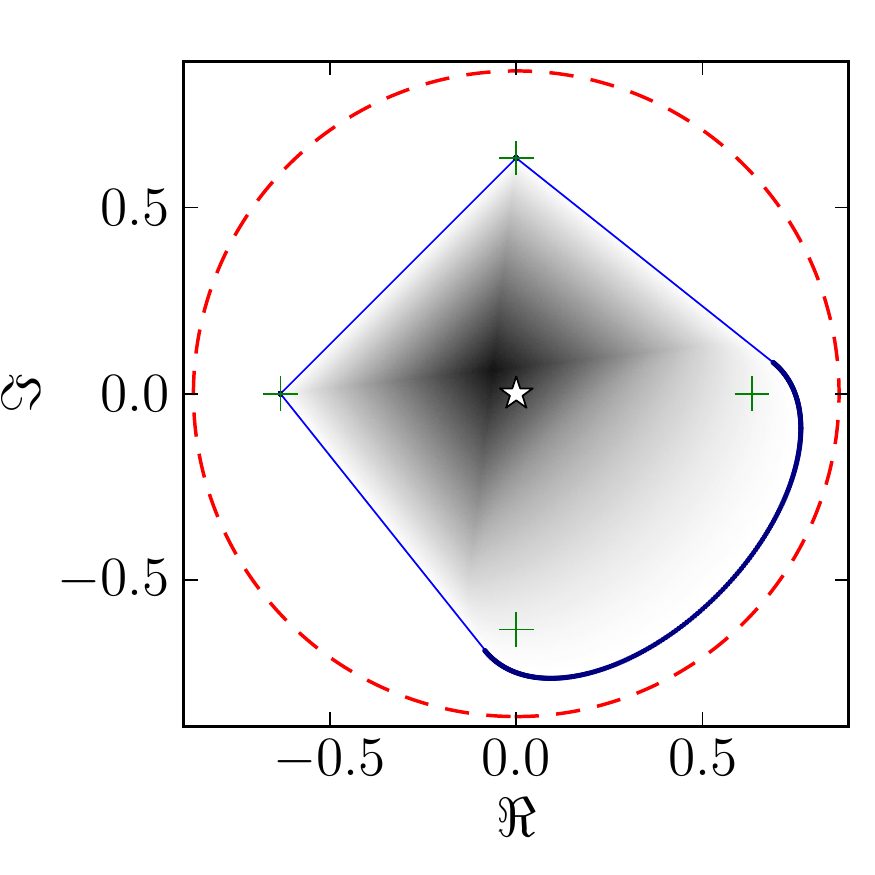}}
\subfigure[$A^{(4)}_5$]{\includegraphics[width=\picwidth]{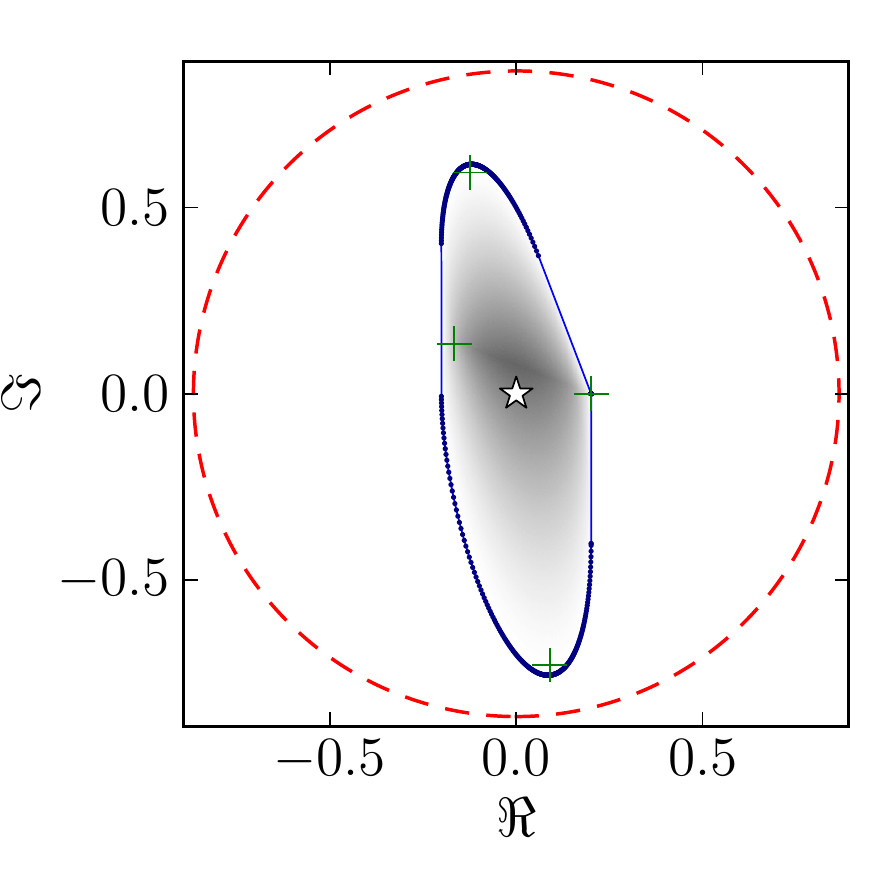}}
\subfigure[$A^{(4)}_6$]{\includegraphics[width=\picwidth]{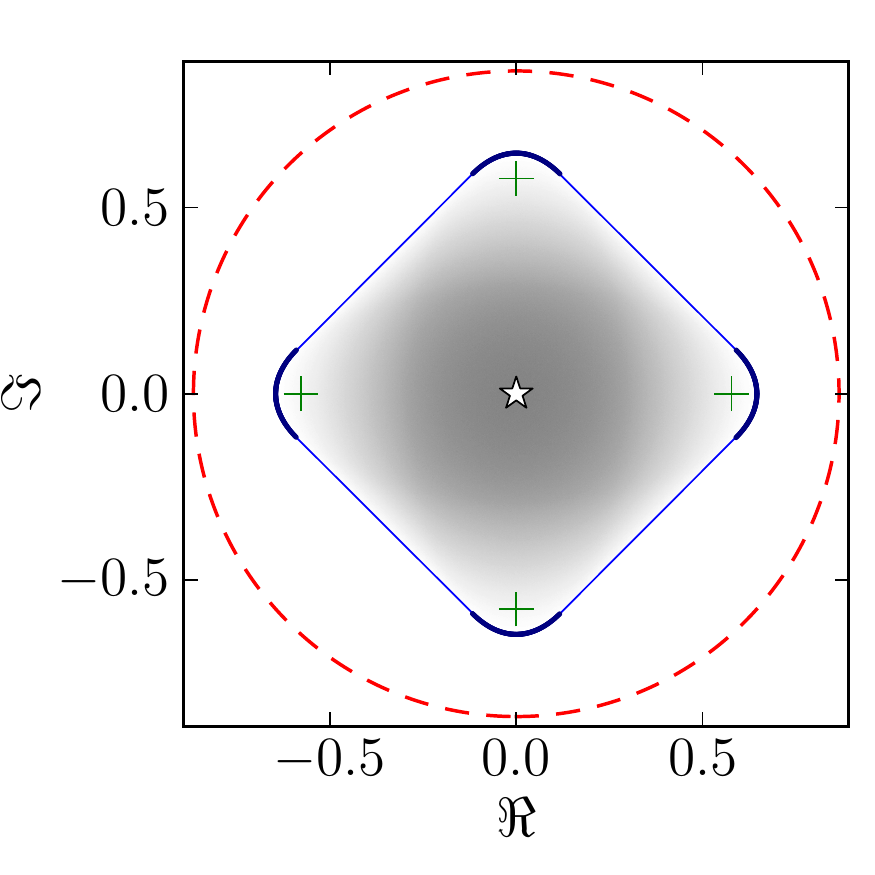}}
\subfigure[$A^{(4)}_7$]{\includegraphics[width=\picwidth]{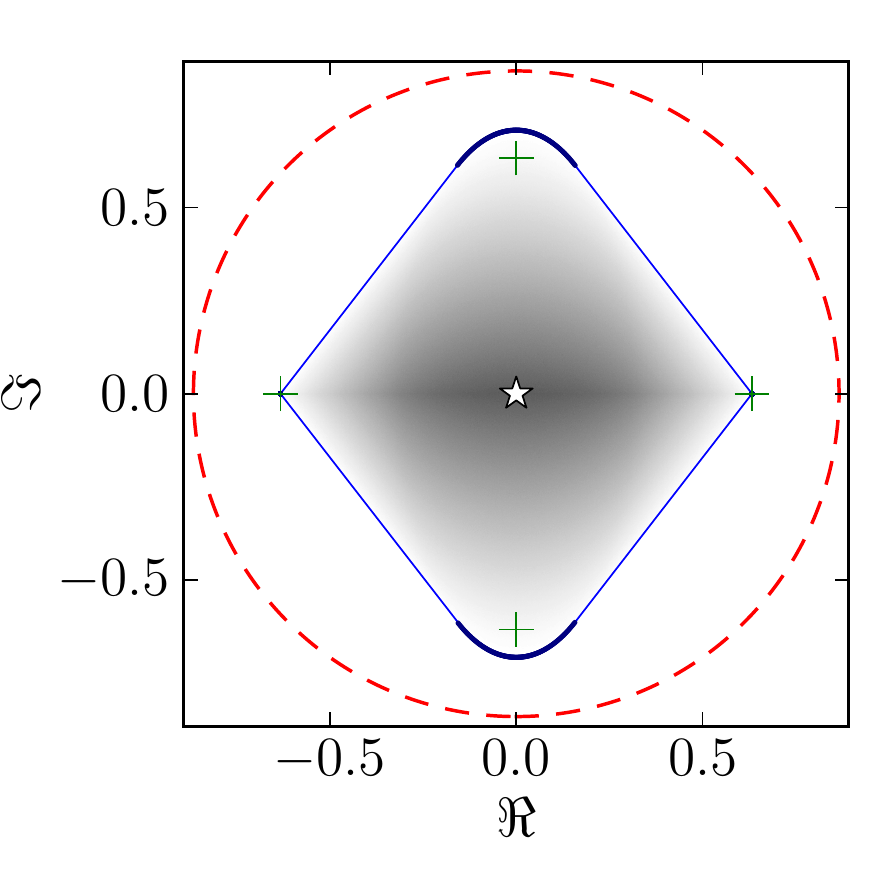}}
\subfigure[$A^{(4)}_8$]{\includegraphics[width=\picwidth]{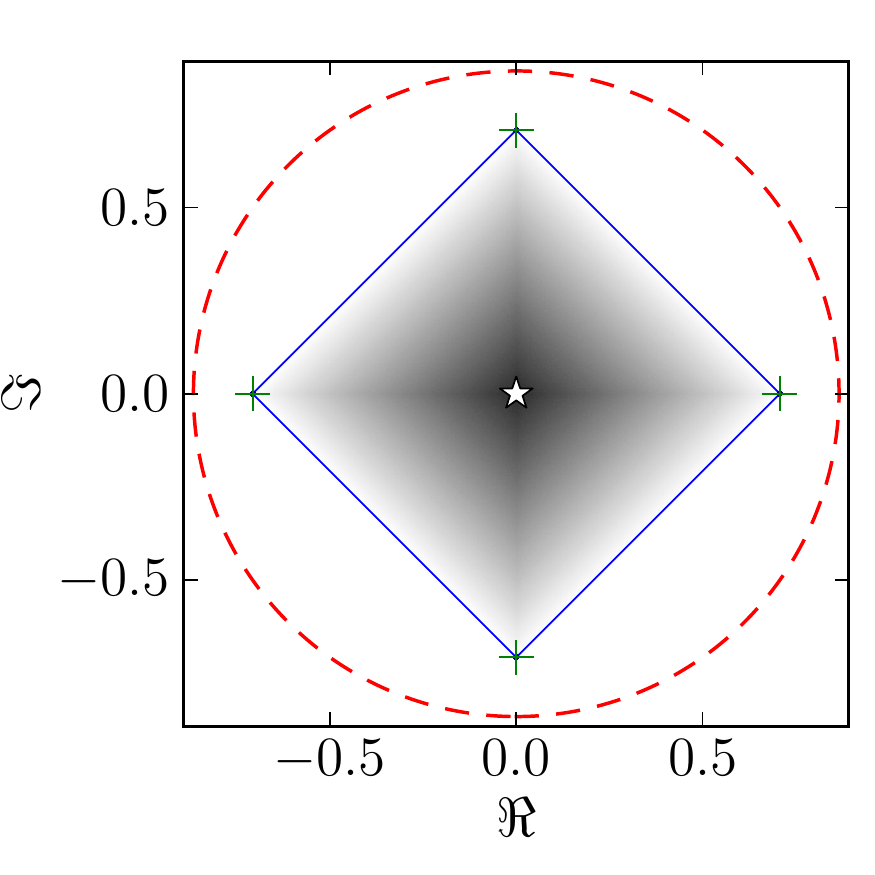}}
\caption{Projections of the set $\Omega_4$ of $N=4$ quantum states 
emerging as numerical shadows of appropriately normalized 
operators of size $4$:
a) a generic matrix $A^{(4)}_0$ with a an oval--like numerical range $W(A)$,
b) $A^{(4)}_1$ with one flat part of the boundary $\partial W$ of the numerical range,
c) $A^{(4)}_{2}$ being a simple sum $3\oplus 1$  with two flat parts of $\partial W$,
d) $A^{(4)}_{3}$ -- a simple sum $2\oplus 2$  with two flat parts of $\partial W$,
e) $A^{(4)}_4$ three flat parts  of $\partial W$ 
        connected with corners and one oval--like part,
f) $A^{(4)}_5$ three flat parts of $\partial W$ with only one corner and two oval--like parts,
g) $A^{(4)}_6$ a simple sum $2\oplus 2$, with four flat parts of $\partial W$, 
h) $A^{(4)}_7$ pair of flat parts of $\partial W$
            connected with a corner connected with two oval--like parts,
i) a diagonal normal matrix $A^{(4)}_8$ with numerical range $W$
  equal to the convex hull of eigenvalues denoted by $(+)$. 
 All plots are done for matrices translated in such a way that their trace
 $(\star)$ is equal to zero and suitably rescaled.
}
\label{fig:shad4}
\end{center}
\end{figure}
Numerical shadows of these representative of each class of $N=4$ matrices are
shown in \fref{fig:shad4}. The pictures can be interpreted as projections of the
$6$-dimensional complex projective space ${\mathbbm C}P^{3}$ onto a plane.
Making use of formula (\ref{alpha}) we find that the normalization constant for
the last example $A^{(4)}_8$ reads $a_8=1/\sqrt{2}$. Thus the diameter of the
shadow, $2a_4=\sqrt{2}$, coincides in this case with the Hilbert-Schmidt distance
between any two orthogonal pure states in ${\cal Q}_4$. The dashed circle of
radius $R_4=\sqrt{3}/2$ represents projection of the sphere into which the set
${\cal Q}_4$ can be inscribed.

%%%%%%%%%%%%%%%%%%%%%%%%%%%%%%%%%%%%%%%%%%%%%%%%%%%%%%%%%%%%%%%%%%%%%%%%%%%%%%%%
\section{Unitary dynamics projected inside the numerical shadow}\label{sec:dynamics} 
%%%%%%%%%%%%%%%%%%%%%%%%%%%%%%%%%%%%%%%%%%%%%%%%%%%%%%%%%%%%%%%%%%%%%%%%%%%%%%%%
As the numerical range and the numerical shadow give us an opportunity to
observe the structure of the space of quantum states, it is possible to apply
these tools to investigate quantum dynamics. A unitary time evolution of a quantum
system is governed by the Hamiltionian operator $H$ (i.e. a self-adjoint
operator representing the total energy of the system),
which leads to  $U(t) =\exp(-\mathrm{i} H t)$.

Note that eigenvalues of the Hamiltonian determine the cyclicity of the
trajectory. The trajectory is periodic iff eigenvalues of the Hamiltonian are
commensurable. The period in this case is given by least
common multiple of the eigenvalues.

Let us consider a three-level system (qutrit). For concreteness, we
choose the Hamiltionian
\begin{equation}
\label{eqn:hamiltonian}
H=
\left[
\begin{array}{ccc}
 -1 & -1-\mathrm{i} & 1 \\
 -1+\mathrm{i} & 0 & 1+\mathrm{i} \\
 1 & 1-\mathrm{i} & 1
\end{array}
\right]
\end{equation}
and select an initial pure state of the system as
$|\psi(0)\rangle=|0\rangle\in {\cal H}_3$. 
The state of the system at some specific time
$t$ is described by the transformed state $|\psi(t)\rangle=U(t)|\psi(0)\rangle$.

\begin{figure}[h!]
\begin{center}
\subfigure[$A_1$]{\includegraphics[width=\picwidth]{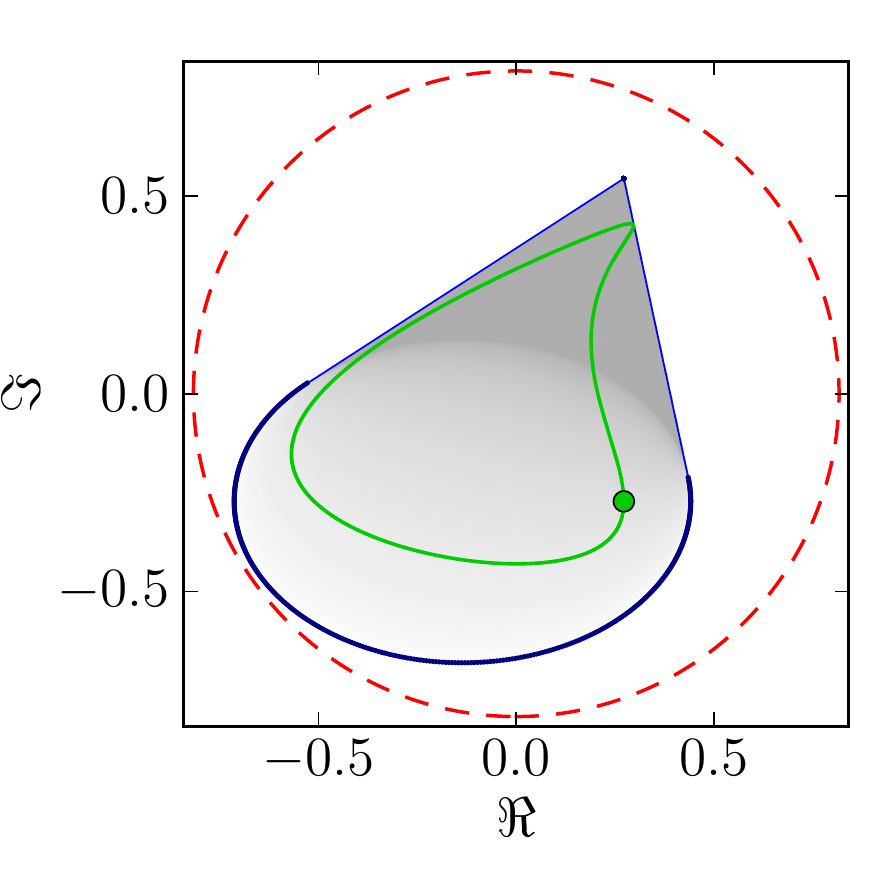}}
\subfigure[$A_2$]{\includegraphics[width=\picwidth]{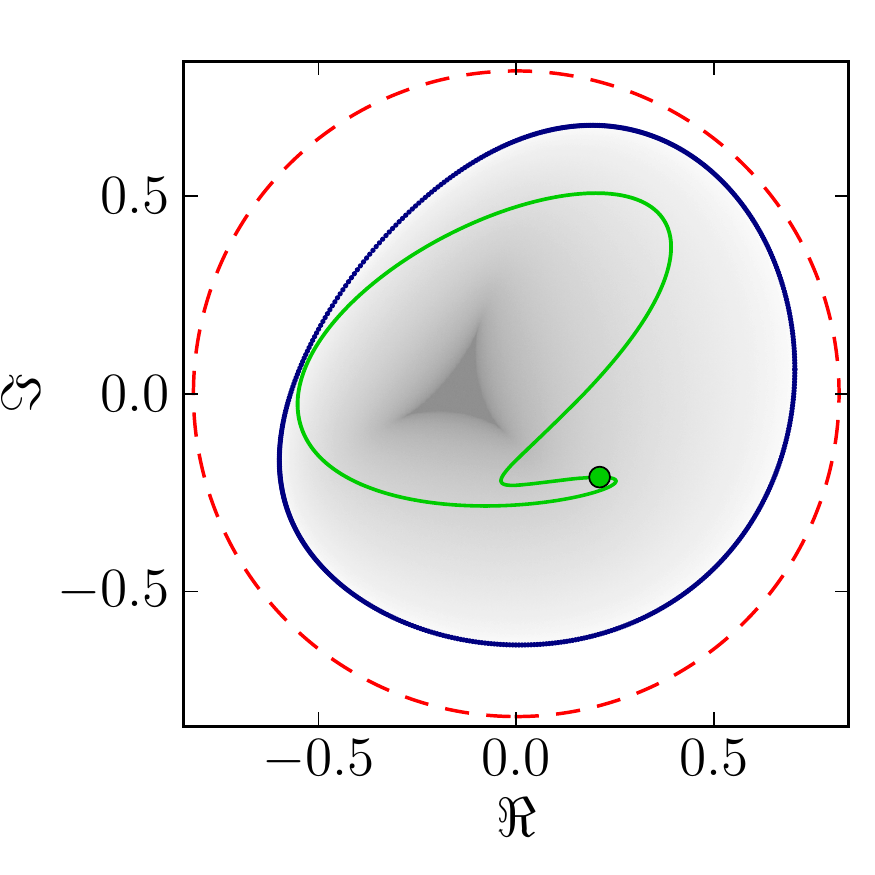}}
\subfigure[$A_3$]{\includegraphics[width=\picwidth]{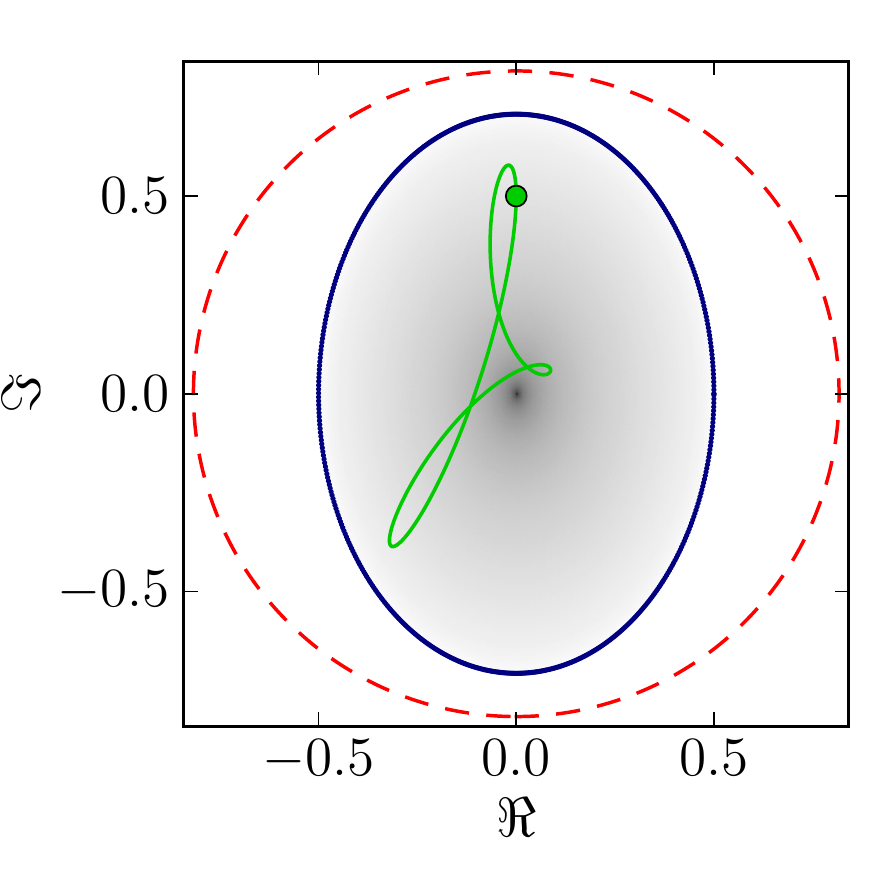}}
\caption{Shadow of the operators $A_1$, $A_2$ and $A_3$ of order three serves as
a background for the trajectory representing the unitary dynamics defined by
$U=\exp(-\mathrm{i}Ht)$ with $H$ given by \eref{eqn:hamiltonian} with the initial state
$|\phi(0)\rangle$ (marked by circle in the picture).
All plots are done for matrices translated in such a way that their trace
is equal to zero and suitably rescaled.
}
\label{fig:dynam1}
\end{center}
\end{figure}

In order to use the numerical shadow to study the time evolution of the system,
one needs to choose an arbitrary $3\times3$ non-hermitian matrix. 
To get some information on the dynamics in the space of pure states of a qutrit
and to observe it from different points of view, 
we selected the following matrices
\begin{equation}
A_1=
\left[
\begin{array}{ccc}
0&0&1\\
0&\mathrm{i}&0\\
0&0&-1
\end{array}
\right],
A_2=
\left[
\begin{array}{ccc}
0&1&1\\
0&\mathrm{i}&1\\
0&0&-1
\end{array}
\right],
A_3=
\left[
\begin{array}{ccc}
\mathrm{i}&0&2\\
0&0&0\\
0&0&-\mathrm{i}
\end{array}
\right].
\label{matdyn}
\end{equation}
For each of these auxiliary matrices, the quantum dynamics can be now visualized as a
trajectory in the complex plane defined by a parametric equation,
\begin{equation} z(t) := \langle \psi(t)|A|\psi(t)\rangle = \langle
\psi(0)|A'|\psi(0)\rangle , \label{unitdyn} \end{equation} where the unitarily
transformed matrix reads $A'=U^*AU=e^{\mathrm{i}Ht}Ae^{-\mathrm{i}Ht}$.
The time evolution of the initial state $|\psi(0)\rangle=(1,0,0)^T\in {\cal H}_3$. 
generated by the Hamiltonian (\ref{eqn:hamiltonian})
is shown in Fig. \ref{fig:dynam1}
from three different perspectives determined by matrices (\ref{matdyn}).

%%%%%%%%%%%%%%%%%%%%%%%%%%%%%%%%%%%%%%%%%%%%%%%%%%%%%%%%%%%%%%%%%%%%%%%%%%%%%%%%
\subsection{Identical trajectories}
%%%%%%%%%%%%%%%%%%%%%%%%%%%%%%%%%%%%%%%%%%%%%%%%%%%%%%%%%%%%%%%%%%%%%%%%%%%%%%%%

For given matrix $A$ and hermitian matrix $H$ a unitary time evolution induces 
a path in the numerical range $\Lambda_A$ given by 
\begin{equation}
\xi^* e^{- \mathrm{i} H t} A e^{ \mathrm{i} H t} \xi, \mathrm{\ for\ given\ starting\ point\ } \xi.
\end{equation}
In the mixed state scenario the trajectory is given by 
\begin{equation}
\tr \rho e^{- \mathrm{i} H t} A e^{ \mathrm{i} H t}, \mathrm{\ for\ given\ starting\ point\ } \rho.
\end{equation}
The question one may pose is: under what conditions for two different starting 
points $\rho_0$ and $\rho_1$ trajectories on numerical range of $A$ are 
identical:
\begin{equation}
 \tr \rho_0 e^{- \mathrm{i} H t} A e^{ \mathrm{i} H t} = \tr \rho_1 e^{- \mathrm{i} H t} A e^{ \mathrm{i} H t}, \ t 
\in \mathbb{R}.
\end{equation}
To convince yourself that such a situation may occur consider the daily rotation
of the earth around its axis. Choosing two initial points at the same meridian
on opposite sides of the equator (say close to Cairo and Durban in Africa), we
see that the trajectories they generate after projecting onto the equatorial
plane do coincide. This is because the dynamics, both initial points and the
kind of the projection, are chosen in a special way and satisfy certain
constraints. To characterize these constrains in a general setting we start with
the following definitions.

For given matrix $A$, let
$X_A = \{B \in M_N^H(\mathbb{C}): \tr B = 0, \ \tr BA = 0 \}$. 
We also define the set $H_A$
\begin{equation}
 H_A = \{ H\in M_N^H(\mathbb{C}): \forall t>0, B \in X_A \mathrm{\ we\ have } 
\ \mathrm{Ad}_{e^{\mathrm{i} H t}} (B) \in X_A \},
\end{equation}
where $\mathrm{Ad}$ is the adjoint mapping given by $\mathrm{Ad}_C(B) = C B C^{-1}$.

Now we can state fact concerning identical trajectories.
\begin{lemma}
Trajectories 
$\tr \rho_0 e^{- \mathrm{i} H t} A e^{ \mathrm{i} H t}$ and $\tr \rho_1 e^{- \mathrm{i} H t} A e^{ \mathrm{i} H t}$ for 
$t \in \mathbb{R}$ are identical if and only if $\rho_0 - \rho_1 \in X_A$ and $H \in H_A$.
\end{lemma}

The definition of $H_A$ is somehow complicated, here we put reasoning which 
presents it in a simpler form.
We have following property $\mathrm{Ad}_{C} = e^{\mathrm{ad}_{C}}$,
where $\mathrm{ad}_{C}(B)  = [C,B]$ (see \emph{e.g.}~\cite{hall2003lie}). Using 
above we can state the following lemma.
\begin{lemma}
Hermitian matrix $H$ is an element of $H_A$, if and only if, for all $B \in X_A$ we have 
$\mathrm{ad}_{\mathrm{i} H} (B)\in X_A$.
\end{lemma}
\proof
If $\mathrm{ad}_{\mathrm{i} H} (B)\in X_A$ for all $B \in X_A$, then by iterating we have
that  $\mathrm{ad}_{\mathrm{i} H}^k (B)\in X_A$ for $k=0,1, \dots$. Since $X_A$ is a 
linear space we obtain, that 
\begin{equation}
 \mathrm{Ad}_{e^{\mathrm{i} t H}} (B) = \sum_{k=0}^{\infty} \frac{t^k}{k!} 
 \mathrm{ad}_{\mathrm{i} H}^k(B) \in X_A. 
\end{equation}
On the other hand if $\mathrm{Ad}_{e^{\mathrm{i} t H}} (B) \in X_A$, then using the fact that
\begin{equation}
 \mathrm{i} [H,B] = \lim_{t \to 0} \frac{e^{\mathrm{i} H t} B e^{-\mathrm{i} H t} - B}{t},
\end{equation}
and the continuity of the function $X \mapsto \tr XA$ we obtain the result.
\halmos

Note, that the condition $\mathrm{i} [H,B] \in X_A$ can be stated as $\tr H [A,B] = 0$, 
this follows from the cyclicity of trace.
The linear space $X_A$ is a real $N^2 - 1 - d(A)$ dimensional space, where $d(A) = \dim(\{\Re(A), \Im(A)\})$,
where $\Re(A) = \frac{1}{2}(A+ A^{*})$ and $\Im(A) = \frac{1}{2 \mathrm{i}}(A- A^{*})$.
Thus in the generic case $X_A$ has dimension $N^2 - 3$.
The set $H_A$ forms a real subspace of hermitian matrices orthogonal to sum of 
two real subspaces 
($\mathrm{ad}_{\mathrm{i} \Re(A)}(X_A)$ and $\mathrm{ad}_{\mathrm{i} \Im(A)}(X_A))$),
\begin{equation}
 H_A = (\mathrm{ad}_{\mathrm{i} \Re(A)}(X_A) \cup \mathrm{ad}_{\mathrm{i} \Im(A)}(X_A))^\bot,
\end{equation}
where $^\bot$ denotes the orthogonal component in the real space of hermitian matrices.

%%%%%%%%%%%%%%%%%%%%%%%%%%%%%%%%%%%%%%%%%%%%%%%%%%%%%%%%%%%%%%%%%%%%%%%%%%%%%%%%
\section{Mixed states numerical shadow}\label{sec:mixed-shadow} 
%%%%%%%%%%%%%%%%%%%%%%%%%%%%%%%%%%%%%%%%%%%%%%%%%%%%%%%%%%%%%%%%%%%%%%%%%%%%%%%%

The standard numerical shadow (\ref{shadow}) of matrix $A$ is
defined by choosing randomly a pure state $\rho=|\psi\rangle \langle \psi|$
with respect to the unitarily invariant, natural measure on the set of pure states, 
and taking the expectation value ${\rm Tr} A \rho$. However, one may also consider
an expression analogous to (\ref{range2}) and use it with a different measure
in the set ${\cal Q}_N$ of mixed states. More precisely,
we introduce the {\sl mixed states numerical shadow} of $A$ with respect to a measure $\mu$,
\begin{equation}
P^{\mu}_A(z)  :=  \int_{{\cal Q}_N} {\rm d} \mu(\rho) 
 \delta\Bigl( z- {\rm Tr} \rho A \Bigr)   .
\label{mishadow}  
\end{equation}
The measure $\mu$ defined on the set ${\cal Q}_N$ of mixed states
of size $N$ is supposed to be unitarily invariant.
For instance, we will use the family of {\sl induced measures} $\mu_K$ 
obtained by taking a random pure state  $|\xi\rangle \in {\cal H}_N \otimes {\cal H}_K$ 
and generating a mixed state by partial trace over the $K$--dimensional subsystem,
$\rho={\rm Tr}_K |\xi\rangle \langle \xi|$.
Since the pure states $|\xi\rangle$ are generated randomly 
the unitary matrices determining the eigenvectors of $\rho$ 
are distributed according to the Haar measure on $U(N)$.
The probability distribution of the eigenvalues $\lambda_i$ 
of the random mixed state $\rho$ of size $N$ obtained in this way reads 
\begin{equation} 
\label{eqn:joint-probab-induced}
 P_{N,K}(\lambda) = C_{N,K} \; \delta \left(1-\sum_{i=1}^N \lambda_i \right) \; 
  \prod_{i=1}^N \lambda_i^{K-N} \prod_{i<j}(\lambda_i - \lambda_j)^2 \; .
\end{equation}
It is assumed here that $K\ge N$ 
and the normalization constants $C_{N,K}$ are given in \cite{ZSo01}.
In the symmetric case, $K=N$, the above formula simplifies
and the measure $\mu_N$ coincides with the flat Hilbert--Schmidt measure,
induced by the metric (\ref{hsdist}).
In the opposite case $K < N$, the joint probability density function is given by 
(\ref{eqn:joint-probab-induced}) with exchanged parameters $N \leftrightarrow K$.

Consider now a pure state $|\xi\rangle$  on the bi-partite $N \times K$ system.
It can be represented in its  Schmidt decomposition \cite{BZ06},
\begin{equation}
 |\xi\rangle  = \sum_{i=1}^{\min\{ N, K \}} \sqrt{\lambda_i} \; |e_i\rangle \otimes |f_i\rangle,
\end{equation}
where $\{|e_i\rangle \}_{i=1}^N$ is an orthonormal basis in  ${\cal H}_N$ while 
$\{|f_i\rangle \}_{i=1}^K$ is an orthonormal basis of ${\cal H}_K$.
Taking a partial trace of the projector 
$|\xi\rangle \langle \xi|$ over the $K$ dimensional system
we see that the spectrum of the resulting mixed state $\rho$
coincides with the set of the Schmidt coefficients $\{ \lambda_i\}$
of the pure state $|\xi\rangle$.
Thus formula (\ref{eqn:joint-probab-induced})
describes  the distribution of the Schmidt coefficients of a pure state $|\xi\rangle$ 
drawn randomly according to the uniform distribution on the sphere $S^{NK -1}$.
By construction of the Schmidt decomposition of a random state $|\xi\rangle$, 
the vectors $|e_i\rangle$ and $|f_i\rangle$ can be considered as
columns of unitary matrix in $U(N)$ and $U(K)$ respectively, distributed 
according to the Haar measure on the unitary group.
A simple  calculation shows that
\begin{eqnarray}
\langle \xi | (A \otimes {\mathbbm 1}_K)|\xi \rangle  &=& 
\sum_{i,j=1}^{\min \{N,K\}} \sqrt{\lambda_i \lambda_j} \langle e_i | A | e_j\rangle 
\langle f_i |f_j\rangle  \nonumber \\
&=&
\sum_{i=1}^{\min\{N,K\}} \lambda_i  \langle e_i | A | e_i\rangle 
=
\sum_{i=1}^{\min\{N,K\}} \lambda_i (U^{\dagger} A U)_{i,i},
\end{eqnarray}
where $U$ is a unitary matrix distributed according to the Haar measure on $U(N)$.

\begin{figure}[ht!]
\begin{center}
 \includegraphics[width=0.7\textwidth]{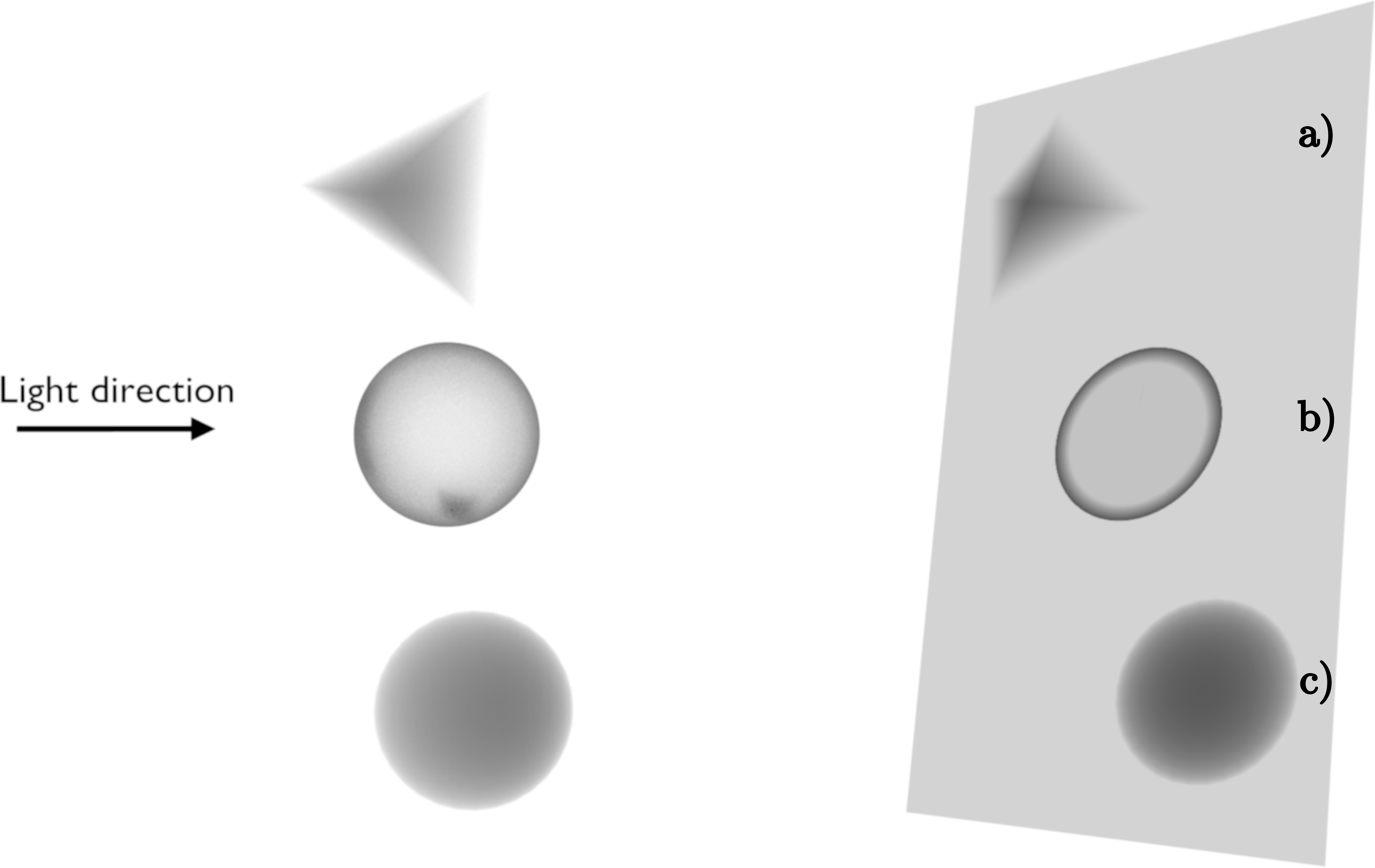}
\caption{
Sketch of a projections onto a two-plane
of 
a) the set ${\cal C}_4=\Delta_3$ of 
$N=4$ classical states onto the quadrangle formed
by the numerical range $W$ of a normal matrix $A$ of order $4$;
b) the set $\Omega_2=S^2$ of one--qubit pure quantum states
onto a disk formed by the numerical range
of a non--normal Jordan matrix $J_2$ of order $2$;
c) a mixed--states numerical shadow of $J_2$,
 corresponding to the projection of the full $3$-dimensional Bloch ball onto a plane,
 is equal to the standard, pure states shadow of an extended
 matrix $J_2 \otimes {\mathbbm 1}_2$.
The picture is plotted using perspective.
}
\label{fig:shadjar}
\end{center}
\end{figure}

These considerations imply that 
the shadow of $A \otimes \mathbbm{1}_K$ is a mixture of diagonal elements of $A$ 
in a random basis, given by the sum
$\sum_{i=1}^{\min\{N,K\}} \lambda_i (U^{\dagger} A U)_{i,i}$.
As before $U$ stands for a random unitary matrix of size $N$ 
while the joint probability distribution function of the 
coefficients $\lambda_i$ is given by (\ref{eqn:joint-probab-induced}).
This proves that the mixed states shadow of $A$ with respect to the induced measure
$\mu_K$ coincides with the standard numerical 
shadow of the extended operator $A\otimes {\mathbbm 1}_K$,
\begin{equation} 
\label{mixed-pure}
 P^{\mu_K}_{A}(z) = P_{A\otimes {\mathbbm 1}_K} (z) =
 P_{{\mathbbm 1}_K \otimes A} (z) .
\end{equation}
The last equality follows from the property (\ref{prod}).
In the most important case, $K=N$, the induced measure $\mu_N$ is equivalent to
the Euclidean (flat) measure in ${\mathbbm R}^{N^2-1}$, corresponding to the
Hilbert-Schmidt distance (\ref{hsdist}). Thus the projection of the 'full' set
${\cal Q}_N$ of mixed quantum states on the plane determined by a given matrix
$A$ of size $N$ is equivalent to the standard shadow of an extended operator $A
\otimes {\mathbbm 1}_N$. In the case of $N=2$ this is visualized in
\fref{fig:shadjar}c), in which the shadow of the full Bloch ball ${\cal Q}_2$
can be compared with the shadow of the hollow Bloch sphere $\Omega_2=S^2$,
displayed in \fref{fig:shadjar}b).

Note that for $K=1$ the induced measure $\mu_1$  is supported on the set 
$\Omega_N$ of pure states only and
coincides with the Fubini--Study measure, so formula 
(\ref{mishadow}) with $\mu=\mu_1$ reduces to the standard 
definition (\ref{shadow}) of the pure states numerical shadow.

%%%%%%%%%%%%%%%%%%%%%%%%%%%%%%%%%%%%%%%%%%%%%%%%%%%%%%%%%%%%%%%%%%%%%%%%%%%%%%%%
\section {Large $N$ limit and random matrices}\label{sec:large-n}
%%%%%%%%%%%%%%%%%%%%%%%%%%%%%%%%%%%%%%%%%%%%%%%%%%%%%%%%%%%%%%%%%%%%%%%%%%%%%%%%
It is instructive to analyse the numerical shadow of a random matrix in the
limit of large matrix dimension $N$. Let us consider two cases of the problem:
the shadow of a random density matrix $\sigma$ generated according to the
induced measures \cite{ZSo01} and the shadow of random unitary matrix $U$
distributed with respect to the Haar measure on $U(N)$.

The numerical shadow of random matrix which is distributed with unitarily invariant measure 
is related to the distribution of its arbitrary diagonal element in a fixed basis. 
In this section  we consider the measures induced by partial trace
and the Haar measure on the unitary group, which are unitarily invariant.
Let us start with the following

\begin{lemma}
\label{lemma-unit-inv}
%{\bf Lemma 7.}
%{\sl 
Let $A$ be a random square matrix of order $N$ distributed according to a unitarily invariant 
measure. Let $\ket{x}$ be a random pure state of size $N$ generated
according to the Fubini-Study measure on  $\Omega_N={\mathbbm C}P^{N-1}$.
Then the expectation value has the same distribution as the matrix element $A_{1,1}$
%}
\begin{equation}
  P\left( \langle x | A |x \rangle \right) =  P\left( A_{1,1} \right) .
\end{equation}
\end{lemma}
%\proof

\proof Since $|x\rangle $ is a random pure state, thus $|x\rangle \sim U |0\rangle$,
where $|0\rangle$ is an arbitrary fixed state while $U$ is a random unitary
matrix of size $N$. Now we write 
\begin{equation}
 P\left( \langle x | A |x \rangle  \right) 
 = P\left( \langle 0 | U^{\dagger} A U | 0\rangle \right) =  P\left( A_{1,1} \right).
\end{equation}
The last equality follows from invariance of the distribution $P(A)$
with respect to unitary transformations.\halmos

%%%%%%%%%%%%%%%%%%%%%%%%%%%%%%%%%%%%%%%%%%%%%%%%%%%%%%%%%%%%%%%%%%%%%%%%%%%%%%%%
\subsection{Shadow of random quantum state}
%%%%%%%%%%%%%%%%%%%%%%%%%%%%%%%%%%%%%%%%%%%%%%%%%%%%%%%%%%%%%%%%%%%%%%%%%%%%%%%%
Consider a random density matrix $\sigma$ of size $N$
generated with respect to the Hilbert-Schmidt measure $\mu_{HS}$,
so that its eigenvalues $\lambda_i$ are distributed according to eq. 
(\ref{eqn:joint-probab-induced}) with $K=N$ \cite{ZSo01}.
The diagonal elements of $\sigma$ are of the form 
\begin{equation}
 \sigma_{ii} = \frac{\sum_{j=i}^N (\xi_{ij}^2 + \eta_{ij}^2 )}
{\sum_{j,k=1}^N (\xi_{jk}^2 + \eta_{jk}^2 )},
\end{equation}
where $\xi_{ij}$ and $\eta_{ij}$ are
independent, identically distributed random variables with normal distribution 
$\mathcal{N}(0,1)$.
Basic properties of the Gamma distribution $\Gamma(a,b)$ \cite{F1986}
imply that 
\begin{equation}
 \sigma_{ii} = \frac{G_1} {G_1 + G_2 } ,
\end{equation}
where $G_1$ and $G_2$ are stochastically independent variables 
distributed according to the Gamma distribution
$\Gamma(N,2)$ and $\Gamma(N(N-1),2)$ respectively.
Therefore the diagonal elements of a random matrix $\sigma$
generated according to the measure $\mu_{HS}$
are described by the Beta distribution with parameters $\{N, N(N-1)\}$.

The same reasoning can also be used for a general class of induced measures
(\ref{eqn:joint-probab-induced}) parameterized by the size $K$ of the auxiliary subsystem.
In this case the diagonal elements of a density matrix $\sigma \in {\cal Q}_N$ generated 
with respect to the  measure $\mu_{N,K}$  
are distributed according to the Beta distribution with parameters $\{K,K(N-1)\}$.

Using the above reasoning and Lemma \ref{lemma-unit-inv} we get the following
\begin{lemma}
The numerical shadow of a random matrix $\sigma$ generated with respect to the
induced measure $\mu_{N,K}$ 
is given by the Beta distribution with parameters $\{K,K(N-1)\}$,
which can be expressed in terms of the Beta function, 
\begin{equation}
P_{\sigma}(r)= \frac{1}{B(K, K(N-1))} (1-r)^{K-1} r^{K(N-1)-1},\quad 0\leq r \leq 1.
\end{equation}
\end{lemma}

%%%%%%%%%%%%%%%%%%%%%%%%%%%%%%%%%%%%%%%%%%%%%%%%%%%%%%%%%%%%%%%%%%%%%%%%%%%%%%%%
\subsection{Random unitary matrices}
%%%%%%%%%%%%%%%%%%%%%%%%%%%%%%%%%%%%%%%%%%%%%%%%%%%%%%%%%%%%%%%%%%%%%%%%%%%%%%%%
Let us now consider a random unitary matrix $U$ distributed according to the
Haar measure.
%{\bf Lemma 8.}
%{\sl 
\begin{lemma}
The numerical shadow of a Haar random unitary matrix is supported in  
the unit disk. This distribution is invariant with respect to rotations, 
and $| \bra{x} U \ket{x} |^2$ is distributed according to the Beta distribution
with parameters $\{1,N-1\}$.
% }
\end{lemma}

%{\bf Proof.}
\proof
Random unitary matrix distributed with Haar measure can be generated using the
QR decomposition of matrices pertaining to the Ginibre ensemble \cite{Mezz07}.
 The QR factorization 
can be realized by a Gram-Schmidt orthogonalization procedure.
Then the element $U_{1,1}$ of the generated unitary matrix reads 
\begin{equation}
 U_{1,1} = \frac{A_{1,1}}{\sqrt{\sum_{i=1}^N |A_{i,1}|^2}},
\end{equation}
where $A$ is a nonhermitian random matrix from the Ginibre ensemble.
Therefore
\begin{equation}
|U_{1,1}|^2 = 
\frac{\xi_{1,1}^2 + \eta_{1,1}^2}
{\sum_{i=1}^N (\xi_{i,1}^2 + \eta_{i,1}^2 )}, 
\end{equation}
where $\xi_{ij}$ and $\eta_{ij}$ are independent, identically distributed random
variables with normal distribution $\mathcal{N}(0,1)$. Thus $|U_{1,1}|^2$ has
the Beta distribution with parameters $\{1,N-1\}$ and using Lemma 
\ref{lemma-unit-inv} we arrive at the desired result. $\square$
%\halmos

%%%%%%%%%%%%%%%%%%%%%%%%%%%%%%%%%%%%%%%%%%%%%%%%%%%%%%%%%%%%%%%%%%%%%%%%%%%%%%%%
\section{Concluding remarks}\label{sec:concluding}
%%%%%%%%%%%%%%%%%%%%%%%%%%%%%%%%%%%%%%%%%%%%%%%%%%%%%%%%%%%%%%%%%%%%%%%%%%%%%%%%
Our study may be briefly summarized by the following observation.
The {\sl numerical shadow} of a normal operator acting on ${\cal H}_N$
reflects the structure of the set of (mixed) classical states,
which belong to the probability simplex $\Delta_{N-1}$,
while investigation of numerical shadows of {\sl non--normal operators}
provides information about the set ${\cal Q}_N$
of quantum states of size $N$.

In particular we have shown that the set of orthogonal projections of the
set ${\cal Q}_N$ of density matrices onto a two--plane is equivalent, 
up to shift and rescaling, to the set of all possible numerical ranges $W(A)$
of matrices of order $N$. The numerical shadow of $A$
forms a probability distribution on the plane, supported in $W(A)$,
which corresponds to the 'shadow' of 
the complex projective space ${\mathbbm C}P^{N-1}$ covered uniformly
according to the Fubini-Study measure, and projected onto the plane.
Another probability distribution in $W(A)$ is obtained if
one projects onto this plane entire convex set ${\cal Q}_N$ 
of density matrices. If this set is covered uniformly
with respect to the Hilbert-Schmidt (Euclidean) measure,
an explicit expression for this distribution is derived.
In this way the analysis of numerical ranges and numerical 
shadows of matrices of a fixed size $N$ contributes to our understanding
of the intricate geometry of the set ${\cal Q}_N$ of quantum states~\cite{BZ06}.

The numerical range \cite{KPLRS09} and its generalizations
\cite{SHDHG08,GPMSZ10} found several application in various problems of quantum
information theory. In analogy to the {\sl product numerical range}, defined for
spaces with a tensor product structure \cite{PGMSCZ10}, one can introduce the
numerical shadow restricted to the subset of separable (product) states or
the set of  maximally entangled states \cite{Zy+10}.
 Analysing such restricted numerical shadows for
operators of a composite dimension $NM$ one may thus investigate the geometry of
the selected set of separable (maximally entangled) quantum pure states. Such an
approach is advocated in a forthcoming publication \cite{DGHMPZ12}.

\ack
It is a pleasure to thank G. Auburn and S. Weis for fruitful discussions.
Work by J.~Holbrook was supported in part by an NSERC of Canada research grant.
Work by P.~Gawron and J.A.~Miszczak was supported by the Polish Ministry of
Science and Higher Education under the grant number N519 442339, Z.~Pucha{\l}a
was supported by Polish Ministry of Science and Higher Education under the
project number IP 2010 033 470, while K.~\.Zyczkowski acknowledges support by
the Polish Ministry of Science and Higher Education grant number N202 090239.
Authors would like to thank S. Opozda for his help with the preparation of 3D
models.

%%%%%%%%%%%%%%%%%%%%%%%%%%%%%%%%%%%%%%%%%%%%%%%%%%%%%%%%%%%%%%%%%%%%%%%%%%%%%%%%

\end{document}